\documentclass{aa}
\usepackage{graphicx}
\usepackage{txfonts}
\usepackage{natbib}
\bibpunct{(}{)}{;}{a}{}{,} 
\begin{document}
%

 \title{The influence of numerical parameters on \\
 tidally triggered bar formation}

    \author{R. F. Gabbasov\inst{1},
            M. A. Rodr\'{\i}guez-Meza\inst{1},
            Jaime Klapp\inst{1}\,\inst{2}
 \and
            Jorge L. Cervantes-Cota\inst{1}}

   \offprints{R. F. Gabbasov}
   \institute{Depto. de F\'{\i}sica, Instituto Nacional de Investigaciones
Nucleares, Apdo. Postal 18-1027, M\'{e}xico D.F. 11801, M\'{e}xico\\
              \email{ruslan@nuclear.inin.mx}
              \and
             Universit\"at Konstanz, Fachbereich Physik, Fach M568, D-78457 Konstanz, Germany\\
              }

   \date{Received September 15, 1996; accepted March 16, 1997}


\titlerunning{The influence of numerical parameters on bar formation}
\authorrunning{Gabbasov et al.}

\label{firstpage}

\abstract{
 The joint influence of numerical parameters such
 as the number of particles $N$, the gravitational softening length
 $\varepsilon$ and the time-step $\Delta t$ is investigated in the
 context of galaxy simulations.
 For isolated galaxy models we have performed a convergence study
 and estimated the numerical parameters ranges for which the relaxed
 models do not deviate significantly from its initial
 configuration.
By fixing $N$, we calculate the range of the mean interparticle
separation $\lambda(r)$ along the disc radius. Uniformly spaced
values of $\lambda$ are used as $\varepsilon$ in numerical tests
of disc heating. We have found that in the simulations with
$N=1\,310\,720$ particles $\lambda$ varies by a factor of $6$, and
the corresponding final Toomre's parameters $Q$ change by only
about $5$ per cent. By decreasing $N$, the $\lambda$ and $Q$
ranges broaden. Large $\varepsilon$ and small $N$ cause an earlier
bar formation. In addition, the numerical experiments indicate,
that for a given set of parameters the disc heating is smaller
with the Plummer softening than with the spline softening.
 For galaxy collision models we have studied the influence of the
 selected numerical parameters on the formation of tidally triggered
 bars in galactic discs and their properties, such as their dimensions,
 shape, amplitude and rotational velocity.
 Numerical simulations indicate that the properties of the formed bars
 strongly depend upon the selection of $N$ and $\varepsilon$.
 Large values of the gravitational
 softening parameter and a small number of particles results in the
 rapid formation of a well defined, slowly rotating bar.
 On the other hand, small values of $\varepsilon$ produce a small,
 rapidly rotating disc with tightly wound spiral arms, and subsequently
 a weak bar emerges.
We have found that by increasing $N$, the bar properties converge
and the effect of the softening parameter diminishes. Finally, in
some cases short spiral arms are observed at the ends of the bar
that change periodically from trailing to leading and vice-versa
-- the wiggle.

\keywords{galaxy interaction -- numerical simulation -- bar
formation}}

 \maketitle

\section{Introduction \label{intro}}
One of the long-standing issues in $N$-body simulations is how a
specific selection of numerical parameters influence an outcome.
In astrophysical and cosmological simulations there are a number
of these parameters that may play an important role, and therefore
they have to be carefully selected. For instance, it is well known
that collisionless simulations of galaxies are performed with a
number of particles $N$, that is far less than the number of stars
in a typical galaxy, hence, it induces artificial non-uniformities
of the gravitational field. In order to smooth the field and to
avoid the divergence of the gravitational force when the
separation between particles becomes small, close encounters
between particles should be avoided. This is usually done by
softening the force between two particles using some modified form
of the Newtonian gravitational potential. For example, the Plummer
softening is widely used since the first simulations of
\citet{b2}. Other softening algorithms exist that give a better
approximation to the Newtonian force such as, for example, higher
powers of the Plummer softening \citep{b4} or the spline softening
\citep{b27}, among others \citep{Dehnen01}. The degree of the
force softening is defined by the gravitational softening
parameter $\varepsilon$. Small values of $\varepsilon$ cause high
relaxation rates but give a better resolution of the internal
structure of the system. On the other hand, a large softening
reduces the relaxation rate, but errors in the force calculation
increase because of the deviation from the Newtonian force.

The influence of $\varepsilon$ on the force calculation error was
discussed by \citet{b31}, where he found that its optimal choice
is related to the minimum error that is characterized by the
so-called bias and variance. He also gives two empirical
expressions to estimate $\varepsilon$ for the \citet{b55} and
\citet{b28} density profiles. This work was later extended to
another configurations \citep{b64, b4} and to other softening
kernels \citep{Dehnen01}. Although these criteria give
$\varepsilon$ that minimizes the force errors, they have been
derived for rigid Monte-Carlo realizations, each of which, when
evolved in time, will drift to its individual equilibrium state.
Such criteria allow to estimate only the magnitude of
$\varepsilon$ that minimizes the initial perturbation due to the
modification of gravity, but ignores the further evolution. For
this reason it is important to check whether these criteria apply
for dynamical evolving systems such as galaxies.

Two-body relaxation effects and force calculation errors depend on
$\varepsilon$ and $N$, and both of them should be simultaneously
minimized. It is, however, difficult to measure them separately
\citep{b30}, e.g. they manifest themselves as a joint error in the
energy of the particles \citep{b60}.

Another source of error stems from an inadequate choosing of the
integration time-step $\Delta t$, which also results in poor
energy conservation. This latter error is typically influenced by
the maximum acceleration, which is constrained by the softening
parameter. In addition, there are truncation and roundoff errors,
but they are small and can be easily controlled.

The problem of an adequate selection of numerical parameters ($N$,
$\varepsilon$, $\Delta t$) for $N$-body simulations has been
widely discussed in the literature, usually separately. In
particular, concerning the effect of the softening parameter in 2D
galaxy simulations, it was established that in addition to
reduction of relaxation, it plays a stabilizing role, and the
meaning of a given value of the $Q$ parameter strongly depends on
the value of $\varepsilon$ \citep{Miller71, Romeo94}. Indeed,
relaxation can heavily affect the results of numerical studies of
disc stability \citep[e.g,][]{White88}. Criteria for choosing
$\varepsilon$ for such systems, based on dynamical requirements
and type of softening, were discussed by \citet{Romeo94, Romeo97,
Romeo98}.
 A 3D study to estimate the numerical parameters was made by
\citet{b26, b30} in which the influence of the softening parameter
and the number of particles on the force errors were investigated.
Also, there is a rich discussion concerning the choosing of
numerical parameters and its effects on cosmological simulations
and small scale structure formation \citep[e.g.,][]{b63}. On the
other hand, analytical estimations give severe limits on the
reliability of $N$-body calculations. For instance, it was shown
that the orbits of particles diverge exponentially with time from
their original trajectories \citep{Goodman93, b65}.

For galaxy simulations, the above mentioned effects influence any
dynamic process, such as the formation of bars in spirals that we
study in the present work. This problem involves physical as well
as numerical aspects. Since the 70s, several numerical simulations
have analysed this problem and found that cold, rotationally
supported stellar discs are globally unstable to bisymmetric
distortions and quickly evolve into well defined spontaneous bars
\citep[e.g.,][]{b58}. The formation of a bar in unstable disc
galaxies was extensively studied in both two- and
three-dimensional simulations \citep[e.g.,][]{b66, b50, b38, b52}.
It was established that the bar formation can be suppressed by
introducing high enough velocity dispersions \citep{b57}, or by a
massive spherical halo \citep{b56}. The 2D-simulations have shown
that the bar's length is not only defined by physical parameters
but also strongly depends on the magnitude of the softening
parameter \citep{b66, b50}. Furthermore, recent 3D-simulations of
isolated galaxies \citep{b14, b16, b6} have shown that the bar
semi-major axis is more than twice as long as the observed length,
which is close to the exponential length-scale of the disc
\citep{b12}. On the other hand, it was found that bar rotation
velocities were slowed down to less than twice the observed
velocities in just a few Gyrs \citep{b16, b6}. This result
coincides with predictions of \citet{b73} that bars are slowed
down by dynamical friction with a massive halo. However, it was
argued that this slowdown is associated with the low halo spatial
resolution that diminishes the rotational velocity of the formed
bar \citep{b40}. A high resolution simulation with $20$ million
particles \citep{ONeill03} still indicates the bar slowdown,
although it is not so drastic. Moreover, it was found that even in
a stable galactic disc a bar may emerge due to discreteness of the
halo \citep*{b69}, and \citet{b70, Athan03} showed that these
effects are related to the disc's angular momentum redistribution
and resonances of halo particles.

Numerical experiments performed by other authors
 have shown that bars may also form in disc galaxies
 which interact with a companion galaxy \citep*{b33, b23, b36}.
A systematic 2D study of tidally triggered bar formation
\citep{b36} in spiral galaxies has shown that bar formation
depends simultaneously on the shape of the rotational curve, the
disc-to-halo mass ratio, the halo concentration, the strength and
the geometry of the perturbation. These results have been
partially confirmed by self-consistent 3D simulations \citep*{b10,
b67, b74}, but the formed bars are found to be slower than in
isolated models, indicating a possible explanation for the
dichotomy of bar characteristics found by \citet{b12}.

Motivated by the above arguments, in this work we investigate for
an isolated galaxy in equilibrium and for the collision of two
galaxies, the range of the numerical parameters ($N$,
$\varepsilon$, $\Delta t$) that permit us to minimize numerical
artifacts in simulations. In addition, we compare the spline
gravitational softening with the Plummer one. For the collision of
two disc galaxies we study the influence of the numerical
parameters on the formation and characteristics of tidal bars.

This work is organized as follows: In Sect. \ref{BHDic} we
describe the initial conditions of the galaxy model and the way in
which the numerical parameters are computed. We obtain a
particular range for ($N$, $\varepsilon$, $\Delta t$), which in
Sect. \ref{conv-stu} is tested to observe deviations from an
initial configuration. Based on a convergence study we choose the
most suitable values for the numerical parameters. Then, in Sect.
\ref{coll} we analyse the influence of the parameters on the bar
formation after the collision of two spirals. Finally, in Sect.
\ref{discu} we discuss the results and in Sect.\ref{conclusions}
outline our conclusions.
%
\section{Numerical modelling} \label{BHDic}

A self-consistent galaxy model is usually constructed with
particles that are governed by the collisionless Boltzmann
equation coupled to the Poisson equation for the Newtonian
gravitational potential. As mentioned above, the gravitational
potential is softened and, ideally, one should construct a
self-consistent model for the potential. For simplicity, a
standard practice is to use an algorithm based on the exact
Newtonian potential, but this requires to relax the initial galaxy
for several crossing times before it can be used for actual
simulations. Using this prescription, we follow \citet{b8} for the
construction of the initial conditions. In this model the bulge
and the halo are both non-rotating spherically symmetric systems
with isotropic velocity dispersions.

The bulge density profile is given by \citep{b28}:
\begin{equation}
\label{rho_b} \rho_{\mathrm{b}}(r)=\frac{M_{\mathrm{b}}
a_{\mathrm{b}}}{2\pi} \frac{1}{r(r+a_{\mathrm{b}})^3} ,
\end{equation}
and for the halo we use Dehnen's $\gamma=0$ density profile
\citep{b18}:
\begin{equation}
\label{rho_h} \rho_{\mathrm{h}}(r)=\frac{3M_{\mathrm{h}}}{4\pi}
\frac{a_h}{(r+a_{\mathrm{h}})^4} .
\end{equation}
Despite the high concentration of this halo model, it was widely
used by Barnes in his galaxy collision simulations. The disc
profile is assumed exponential \citep{b22}:
\begin{equation}
\label{rho_d} \rho_{\mathrm{d}}(r,z)=\frac{M_{\mathrm{d}}
\alpha^2}{4\pi z_0} e^{-\alpha r} \mbox{sech}^2 \left(
\frac{z}{z_0} \right).
\end{equation}
In equations (\ref{rho_b}-\ref{rho_d}) $M_{\mathrm{b}}$,
$a_{\mathrm{b}}$ and $M_{\mathrm{h}}$, $a_{\mathrm{h}}$ are the
mass and scale-length of the bulge and the halo, respectively, and
$M_{\mathrm{d}}$, $\alpha^{-1}$ and $z_0$ are the mass, the
scale-length and the scale-height of the disc, respectively.

Using the Monte-Carlo technique, the density profiles are
 represented by a system of $N$ equal mass bodies,
 $N$ being the total number of particles.
For reducing two-body relaxation effects and to avoid the mass
segregation \citep[e.g.,][]{FaroukiS94} we take equal mass
particles. The number of particles in each component are assigned
in proportion to their masses:
 $M_{\mathrm{b}}:M_{\mathrm{d}}:M_{\mathrm{h}}= 1:3:16$.
 For each component, the mass distribution is truncated at a radius
 that contains $95$ per cent of its total mass, that would correspond
 to an infinite radius.

\begin{figure}
\begin{center}
\includegraphics [width=84mm]{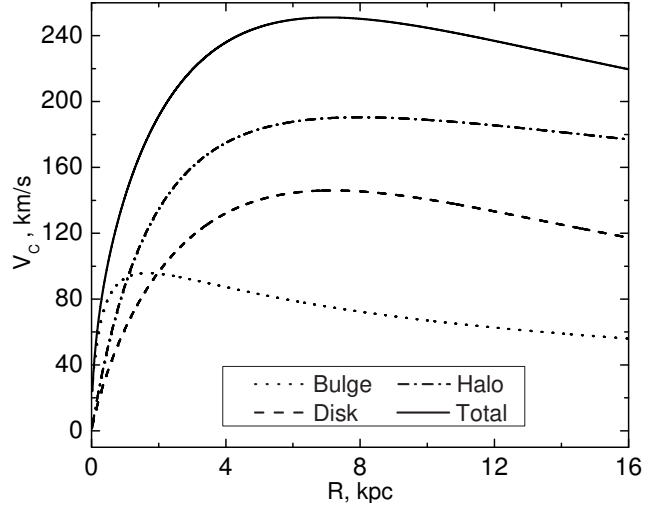}
\caption{Rotation curves of the model. The bulge and halo rotation
curves are shown up to the radius of the disc.} \label{bar_fig1}
\end{center}
\end{figure}

 The disc vertical dispersion $\sigma^2_z(r)$ is found
 from an isothermal sheet equilibrium condition,
 and the radial velocity dispersion is $\sigma^2_r(r)=4\sigma^2_z(r)$.
 The azimuthal velocity dispersion is
 calculated using the epicyclic approximation
 $\sigma^2_{\phi}(r)=\sigma^2_r(r)\kappa^2/4\Omega^2$,
 where $\kappa$ is the epicyclic frequency and $\Omega$ the
 angular velocity. The net rotational velocity of the disc is computed from
 the asymmetric drift equation \citep{b46}.
 The bulge and halo dispersion velocities are
 calculated using equation (\ref{dispersion_r}), see below.
 This equation was
 integrated numerically, including the masses of all components.
 Then, isotropic velocities are drawn from the Gaussian distribution.
 For a detailed description of the initial conditions see
 \citet{b8} and \citet{b29}.
The parameters for our galaxy model are summarized in
Table~\ref{model}, where the columns from left to right give the
mass, the number of particles, the cut-off radius and the length
scale of each component. We use Barnes's model parameters and
system of units. The disc's scale-height is $z_0=0.007$.
 The mass, length and time scales are set to
 $2.2 \cdot 10^{11}\ {\mbox M}_{\sun}=1$, $40\
 {\mbox {kpc}}=1$ and $250\ {\mbox {Myrs}}=1$. In these units,
 the gravitational constant is $G=1$. The half mass radius of the
galaxy is located at $R_{1/2}\approx 11$ kpc.
 The total rotational curve and rotation curves of each component are
 shown in Fig.~\ref{bar_fig1}.
\begin{table}
\caption{Parameters of the galaxy model. The units of mass and
length are $2.2 \cdot 10^{11}\ {\mbox M}_{\sun}$ and $40\
 {\mbox {kpc}}$, respectively.} \label{model}
\begin{tabular}{llrcr} \hline\hline
Component & Mass & Number of & Cutoff & Scale-\\
 & & particles & radius & length\\
\hline
 Bulge & 0.0625 & $0.05 N$ & 1.5 & 0.04168\\
 Disc & 0.1875 & $0.15 N$ & 0.4 & 0.0833\\
 Halo & 1.0 & $0.8 N$ & 6.0 & 0.1\\
\hline
\end{tabular}
\end{table}
%
\subsection{Methods}

For the time evolution we use a hierarchical General Body Smoothed
Particle Hydrodynamics (GBSPH) treecode  written by one of us
(M.A.R.M., see details in {\it www.astro.inin.mx/mar/nagbody})
which is similar to Barnes' treecode \citep{b7}. We use the
Plummer softened point-mass potential, and the forces are computed
up to quadruple terms with a tolerance
 parameter $\theta=0.75$. The equations of
motion are integrated using a second order leap-frog algorithm
with a fixed time-step. The SPH part was switched off.

To verify our results, we also performed several runs using the
serial and parallel versions of the public GADGET code
\citep{b44}. The main difference between the codes is that for the
Newtonian potential, GADGET uses a spline approximation. The
latter calculates for the given Plummer softening the equivalent
spline softening $h=2.8 \varepsilon$. In order to maintain
numerical similarity, we have fixed the time-step to a single
value and set the rest of the parameters equal to those used by
the GBSPH code. Also, the tree structure was updated at each
time-step.
%
\subsection{Selection of the numerical parameters} \label{op-par}

For numerical studies one pursues to have
 a collisionless, stable equilibrium system. But, due to
 discreteness effects and force errors this condition is not
 perfectly fulfilled. If the numerical parameters are accordingly chosen,
 a system that closely matches an initial stable equilibrium during a
 specific time can be achieved.
 Thus, the selection of the parameters is of primary importance
 in any $N$-body simulation.

In spite of the many works on the subject there is still no
suitable criterion to define the appropriate magnitude of
$\varepsilon$ for a given number of particles. Usually,
$\varepsilon$ is chosen in an ad hoc way, or by performing various
runs with different $\varepsilon$ and selecting the one that gives
a better conservation of the total energy and angular momentum
and/or an initial density profile. But this procedure demands a
large amount of computational time and is not always performed. On
the other hand, a common practice to verify the results is to
rerun the simulation with higher $N$, leaving $\varepsilon$ and
$\Delta t$ unchanged.

In fact, given the density profile and the number of particles one
can roughly estimate the required values of the gravitational
softening parameter and the time-step. We now proceed to analyse
the criteria for choosing the numerical parameters in order to
constrain their possible values for the above galaxy model. This
subsection has a preparatory purpose for the numerical simulations
of the next section, that should further confine the parameters.
The set of parameters that minimize the disc heating and total
energy errors will be called ``optimal''.
%
\subsubsection{The number of particles}

It is well known that in collisionless $N$-body simulations, the
number of particles should be as large as possible. With the new
parallel codes it is now possible to simulate galactic systems
with $N \ga 10^6$--$10^7$.

The minimum number of particles $N$ required to sample
 a mass $M$ inside an homogeneous sphere of radius $R$
 can be estimated from the relaxation time for the softened potential
\citep*{b48, b59}:
\begin{equation}
\label{t_relax} \frac{t_{\mathrm{relax}}}{t_{\mathrm{cross}}}=
\frac{N}{8\ln\left(R/ \varepsilon\right)},
\end{equation}
where $t_{\mathrm{cross}}\simeq(G\overline{\rho})^{-1/2}$ is the
crossing time, and $\overline{\rho}$ is the mean density of the
system. Once $\varepsilon$ is defined and the $t_{\mathrm{relax}}$
is demanded to be comparable to the age of the universe, the value
of $N$ can be found. For example, by requiring
$t_{\mathrm{relax}}>10^{10}$ years, the halo model given by
equation (\ref{rho_h}) with $t_{\mathrm{cross}}\approx 1.5\cdot
10^9$ years and $\varepsilon=0.01$ will need a minimum
$N_{\mathrm{min}}\approx 3300$, which is quite a small number.
However, numerical simulations \citep{b59} show that this
condition underestimates the value of the $N$ for centrally
concentrated systems by roughly an order of magnitude.

On the other hand, for realistic galactic disc simulations, the
vertical structure should be resolved, although the general
structure in the plane of the disc can be simulated even when
$\varepsilon>z_0$ \citep{b42}. In order to resolve the vertical
structure of the disc component given by equation (\ref{rho_d}),
one may require $\lambda(R_{1/2}) \le z_0$, where $\lambda$ is the
mean interparticle separation estimated within the disc's
half-mass radius $R_{1/2}$, and  $z_0$ is the vertical scale
height \citep{b26}. For our model this gives
$N_{\mathrm{min}}\approx 10^4$ disc particles. In principle, this
number can be considered as an acceptable minimum for our galaxy
model. However, in order to make a more realistic estimation we
have to further analyse the dependence of $N$ with $\varepsilon$
and $\Delta t$.
%
\subsubsection{The softening parameter}

For a given minimum relaxation time and $N$, the magnitude of
$\varepsilon$ depends on the mass distribution of the gravitating
system. In a series of papers \citet[][ and references
therein]{Romeo98} has analysed the dispersion relation for a 2D
disc and discussed the criteria for physically consistent values
of the softening parameter. It was found that stability and
relaxation impose different requirements on
 $\varepsilon$. Although this fact also have implications for 3D discs,
 the question in this case is much more complicated because the particles' vertical
motion has to be taken into account.

For an homogeneous 3D mass distribution a natural choice for
$\varepsilon$ is to be equal to the mean interparticle separation
$\lambda$. But astrophysical systems are centrally condensed and
the mean interparticle separation is determined by local averages,
hence $\varepsilon$ is a local parameter. In practice, however,
some authors prefer to compute the minimum constant $\varepsilon$
that optimize some force accuracy level \citep{b26}. An
inconvenience arises for very inhomogeneous systems since the
election is not obvious.

There are several popular criteria that are widely used:
\begin{enumerate}
\item \citet{b42} have suggested that the softening parameter of a
gravitating disc can be taken proportional to the radius of
interaction, $r_{\mathrm{int}}$, of the particles in its plane,
which is given by
\begin{equation}
\label{int_rad} r^2_{\mathrm{int}}=\frac{m}{M(r)}r^2,
\end{equation}
where $m$ is the mass of an individual particle and $M(r)$ is the
total mass of the disc inside the radius $r$.
 The field is considered smooth if the distance at which the interaction
 force between a
 test particle and its nearest neighbour is equal to the force between the
 same particle and the rest of the particles in the disc.
 The system is said to be collisionless if the softening parameter is
 chosen such that $\varepsilon > r_{\mathrm{int}}$.

\item \citet{b26} have studied force calculation errors, produced
by small number of particles and the Plummer softening of the
gravitational potential. He showed that errors decrease
 weakly with increasing $N$, and are minimized if $\varepsilon < \lambda$,
 where $\lambda$ is the mean interparticle separation
 evaluated within the half-mass radius $R_{1/2}$.
 For $\varepsilon > \lambda$ the errors increase rapidly.
 The latter is explained, particularly for treecodes, by the fact that
 the expansion of cluster potentials is accurate only for pointlike
 particles, and it fails for highly softened particles \citep{b26}. Then,
 Newton's third law is violated and linear momentum is
 badly conserved. Thus, this criterion can be used to impose an upper
 limit for $\varepsilon$.

\item For a spherical gravitating system the value of
 $\varepsilon$ that minimizes force errors can be estimated
from an empirical relation \citep{b31}:
\begin{equation}
\label{athanassoula_eps} \varepsilon_{\mathrm{a}}=AN^{a},
\end{equation}
where $A$ and $a$ are parameters that depend on particular models.
For example, for a Dehnen's $\gamma=0$ density profile of total
mass $M=1.0$, scale-length $a=0.1$ and the cutoff radius
$R=299.8$, their values are $A=0.12$ and $a=-0.27$, respectively
\citep{b4}. However, as it was mentioned in the introduction, this
criterion only reduces the initial shot noise, but does not
guarantee that the softening chosen in such way will be valid for
a long term evolution.

\end{enumerate}

Thus, combining these criteria, it is possible to restrict the
values of the softening for a gravitating system. Let us consider
a spherical gravitating system of mass $M$ in steady state
equilibrium.
 It can be divided in spherical shells of thickness $\bigtriangleup r$
 whose volumes are
\begin{equation}
\label{dv}
\bigtriangleup V(r,r+\bigtriangleup r)=\int \limits_r^{r+\bigtriangleup r}
 \!\!{\mathrm{d}}^3{\bf r} .
\end{equation}
Given the density profile $\rho({\bf r})$, the mass of the shell is
\begin{equation}
\label{dm} \bigtriangleup M(r,r+\bigtriangleup r)=\int
\limits_r^{r+\bigtriangleup r} \!\!\rho({\bf r})\,
{\mathrm{d}}^3{\bf r} .
\end{equation}
If the system is represented by a Monte Carlo distribution of $N$
particles with equal masses $m=M/N$,
 then the number of particles in that shell is roughly
\begin{equation}
\label{dn}
\bigtriangleup N(r,r+\bigtriangleup r) =
 \frac{\bigtriangleup M(r,r+\bigtriangleup r)}{M}N .
\end{equation}
 The mean interparticle separation in the shell is then given by
\begin{equation}
\label{lambda} \lambda(r,r+\bigtriangleup r) = \left[
\frac{\bigtriangleup V(r,r+\bigtriangleup r)}{\bigtriangleup
 N(r,r+\bigtriangleup r)}\right]^{1/3} .
\end{equation}

\begin{figure}
\begin{center}
\includegraphics [width=84mm]{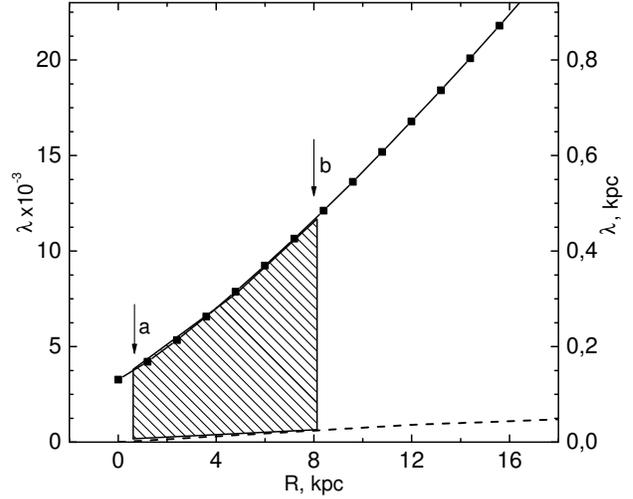}
\caption{Mean interparticle separation as function of radius for
 Dehnen's $\gamma=0$ density profile (upper curve), and the interaction radius
 (lower, dashed curve).
 The curves are plotted only up to $R=18$ kpc in order to show the behavior
 near the centre.
  The arrows mark (a) the value of $\varepsilon_{\mathrm{a}}$ calculated from
 equation (\ref{athanassoula_eps}), and (b) $\lambda$ evaluated within
  the half-mass radius.}
\label{bar_fig2}
\end{center}
\end{figure}

Thus, for each shell a corresponding softening $\varepsilon(r)
\propto \lambda(r)$ can be found. However, one cannot use a
radially varying gravitational softening in the simulations, since
it will lead to energy non-conservation \citep{b42}. To avoid
this, one needs a single value of $\varepsilon$ to be used as the
gravitational softening parameter for the whole system.
 When a value of $\varepsilon$ is estimated at some
radius $R_{\mathrm{res}}$, it defines the radius of the system
within which the structure is poorly resolved and the obtained
results for this region have to be treated with caution.
 Fig.~\ref{bar_fig2} shows $\lambda(r)$ for the above Dehnen's sphere
 represented by $N=262\,144$ particles together with the values of
(a)
 $\varepsilon_{\mathrm{a}} \approx 0.0041$
 and (b) $\lambda (<R_{1/2}) \approx 0.0122$,
  that are marked by arrows.
 Alternatively, one may use the interaction radius
 defined by equation (\ref{int_rad}) as an estimate for $\varepsilon$.
  The interaction radius $r_{\mathrm{int}}$, calculated assuming circular orbits,
 increases linearly with radius from zero to roughly $3.31$ at the cutoff radius.
 Using these criteria the possible values of softening are
confined inside the striped area.

In order to estimate the inferior limit for $\varepsilon$, the
following procedure can be used. The Monte-Carlo particles are
indeed agglomerations of $\sim 10^4$--$10^6$ M$_\odot$ when
compared to real galaxies. Assuming the particles to be Plummer's
spheres with masses $m$ and half-mass radii $r_{\mathrm{h}}$, they
can be treated as point masses if the local mean separation
$\lambda\ge 2r_{\mathrm{h}}$. In order to estimate the value of
$\varepsilon$, we can use the same opening criterion $\theta$ that
is used in treecodes. The structure of the mass distribution is
unresolved when $2r_{\mathrm{h}}/\lambda\le \theta$.
 The softening radius of a Plummer's
sphere is related to $r_{\mathrm{h}}$ as $1.3\varepsilon \approx
r_{\mathrm{h}}$, and thus
\begin{equation} \label{epsilon}
\varepsilon = \theta \lambda/2.6\ .
\end{equation}
If $\lambda$ is known, a rough value of $\varepsilon$ can be
calculated. As shown by \citet{b26}, for treecodes the value of
$\theta \in (0.7-1)$ represents a compromise between accuracy and
performance. However, there is an inconsistency in implementing
the Plummer softening due to force calculation as it would be
between a point mass and a Plummer sphere, instead of between two
point masses \citep{b20}.
 With the above assumption, the latter is weakened, but not completely removed
 because of the infinite extension of the Plummer distribution.
While this inconsistency should be taken into account in
simulations of systems such as globular clusters, some authors
argue it is irrelevant for galaxy simulations
\citep[e.g.][]{Romeo98, Dehnen01}. On the other hand, the study of
\citet{b65} shows that the adequate softening is in the range
$\varepsilon \in [0.2-1]$ times the mean interparticle separation,
in agreement with equation (\ref{epsilon}) and
Fig.~\ref{bar_fig2}.

For each component (bulge, disc and halo) of the galaxy model, the
mean interparticle separation can be easily evaluated as a
function of radius using equations (\ref{dv}-\ref{lambda}), and
$\varepsilon$ from equation (\ref{epsilon}). Once $\varepsilon$ is
defined for each component, it can be used as an input softening
parameter for codes such as GADGET that handle individual
$\varepsilon$ for each component.
 Individual gravitational softenings are preferable in
 simulations, but the total energy is less conserved and Newton's third law
 is strongly violated \citep{b20}. To avoid this problem individual
 softenings should be somehow symmetrized in order that particles
 have the same maximum force.
Besides, the $\varepsilon$ found in this way does not include the
mass distribution of the other components.

If a single value of $\varepsilon$ is used for a whole galaxy
 model, then the question becomes more complicated, and some
 averaged over component value have to be chosen.
 Clearly, the single value of $\varepsilon$ will resolve well only some
 intermediate region of the galaxy model, but at the same time,
 there will be regions subject to numerical heating and poor resolution.

As is shown by \citet{b4} in a compound system the force error
basically stems from the densest component and depends on its mass
fraction. In a galaxy model, high density regions are located at
the centre and in the plane of the disc, while low density regions
are found at the periphery, mostly in the external halo.
 Since we are interested in processes occurring in the disc component,
 this should be well resolved.
 Thus, we compute the particle density over the cylindrical volume
 of the disc, which encompasses all three components of the galaxy:
\begin{equation}
\label{N_cyl_tot} \bigtriangleup N(r,r+\bigtriangleup r)
=\sum_{i=1}^3 \bigtriangleup N_i(r,r+\bigtriangleup r) ,
\end{equation}
with
\begin{equation}
\label{N_cyl_i} \bigtriangleup N_i(r,r+\bigtriangleup r) =
 \frac{\bigtriangleup M_i(r,r+\bigtriangleup r)}{M_i}N_i .
\end{equation}

\begin{figure}
\begin{center}
\includegraphics [width=84mm]{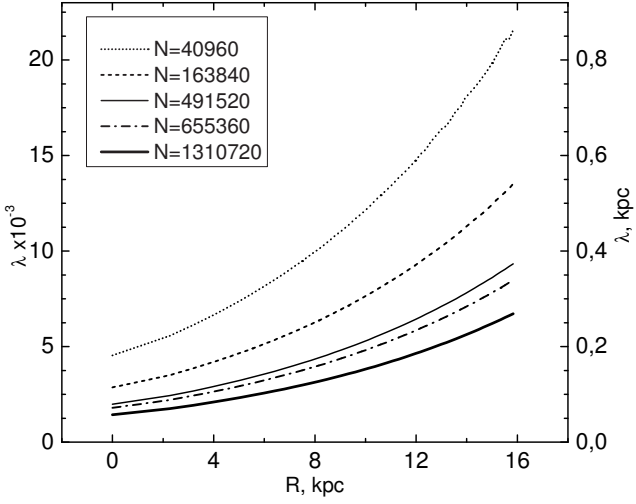}
\caption{Mean interparticle separation as a function of radius for
the model galaxy for different $N$.} \label{bar_fig3}
\end{center}
\end{figure}

 The total $\lambda(r)$ for our galaxy model for different $N$
 is shown in Fig.~\ref{bar_fig3}.
 The volume of the disc was divided in $50$ equal volume concentric annuli
 and the mean separation was calculated in each annulus using equations
(\ref{lambda}) and (\ref{N_cyl_tot}).
It is interesting that with increasing $N$, the variation of
$\lambda$ shallows. Of course, the shape of these curves will
depend on the mass distribution of a particular galaxy model.
%
\subsubsection{The time-step}

The accuracy of an integrator scheme is determined by the
time-step, $\Delta t$, which is in turn constrained by some
criteria, e.g. the Courant condition. If two particles become
close to each other, their acceleration increases. In order to
handle their orbits adequately, the time-step should be reduced.
Most $N$-body codes implement variable time-steps to achieve both
a better efficiency and accuracy. These codes use for each
particle an individual time-step, with criteria based on its local
velocity and/or acceleration, see for example \citet*{b44}.
However, for codes that use a single time-step value for the
evolution of the whole system, the magnitude of $\Delta t$ should
be carefully estimated. Because the maximum acceleration is
constrained by the softening parameter and this is related to the
number of particles, in choosing the time-step one has to consider
the role of both parameters.

In order to find the value of the time-step which is required for
the correct integration of the orbits of particles in the shell,
one may use the Courant condition which requires knowing their
mean velocity of motion. In a spherical system in equilibrium
steady state the velocities of the particles are usually derived
from the Gaussian velocity distribution. The velocity dispersions,
that are functions of the radius, can be obtained by taking the
second momentum of
 the collisionless Boltzmann equation \citep{b46}.
For an isotropic velocity distribution of a spherically symmetric
system, $\sigma_r = \sigma_{\varphi} = \sigma_{\theta}$, with
\begin{equation}
\label{dispersion_r} \sigma^2_r (r) = \frac{1}{\rho(r)}\int
\limits_r^{\infty}\rho(r') \frac{G M(<\!r')}{r'^2}{\mathrm{d}}r',
\end{equation}
where $\rho(r)$ is the density and $M(<\!r')$ is the mass inside
a sphere of radius $r'$. The mean square velocity of particles
in a shell of thickness
 $\Delta r$, is
\begin{equation}
\label{v_mean} <\!v^2 (r,r+\bigtriangleup r)\!> = 3 <\!{\sigma^2_r
 (r, r+\bigtriangleup r)}\!>,
\end{equation}
with
\begin{equation}
\label{sigma_mean} <\!{\sigma^2_r(r, r+\bigtriangleup r)}\!> =
\frac{1}{\bigtriangleup M(r,r+\bigtriangleup r)}\int
\limits_r^{r+\bigtriangleup r}\!\!\sigma^2_r (r)\rho(r)\,
{\mathrm{d}}^3{\bf r}.
\end{equation}
The time required to travel the distance $\lambda$ in that shell is
\begin{equation}
\label{t_mean} <\!\bigtriangleup t (r,r+\bigtriangleup r)\!> =
 \frac{\lambda(r,r+\bigtriangleup r)}
{<\!v (r,r+\bigtriangleup r)\!>}.
\end{equation}
The space resolution is defined by the radius $R_{\mathrm{res}}$,
and the minimum time-step may be taken as the one estimated at
$R_{\mathrm{res}}$. Also, one can average the value of
$<\!\bigtriangleup t (r,r+\bigtriangleup r)\!>$ over the shells,
obtaining the corresponding to the mean velocity of the system.
This average time can be used as an estimate of the time-step for
the orbit integrator. As a consequence, the central region having
too short dynamical times will not be resolved.

For our galaxy model the maximum velocity is determined by the
rotation curve ($\sim 250$ km\,s$^{-1}$) and the velocity averaged
over components varies with radius between $160$ and $120$
km\,s$^{-1}$.

Though in galaxy simulations one uses Newtonian dynamics, the
finite propagation speed of the gravitational interaction should
be included for a realistic self-consist model. If gravity
propagates with the speed of light $c$, then information about a
particle on the opposite side of the system of radius $R=150$ kpc,
will be received after $\Delta t=2R/c\approx 10^6$ years, or $\sim
1/250$ in our units. This implies a restriction on the minimum
allowed time-step within which the information is received by all
particles, and/or the maximum radius of the system that can be
considered in order to maintain the consistency of the simulation.
However, it can be shown that on galactic scales the error
introduced by particles exterior to $c\Delta t$ is not greater
than errors caused by the force approximation. The Newtonian
approach can either be used when a steady state has been reached,
because then the number of particles in a volume element maintains
roughly the same, or when the system is not too extended and
contribution from distant particles can be neglected. For a
non-equilibrium rapidly evolving system, such as a cluster of
galaxies, retarded gravitational potentials should be used.

In addition, the range of $\Delta t$ is restricted from a
numerical point of view, by the requirement of stability of the
integrator algorithm and the required accuracy.

\bigskip
Thus, the expressions given by equations (\ref{lambda}) and
(\ref{t_mean}) permit us to define
 suitable values for the mean interparticle separation and the time-step as
 a function of the system's radius. The advantage is that
 they can be calculated during the construction of the
 model with the number of particles as an input parameter.
It is clear that the selection of a single value for $\varepsilon$
and $\Delta t$ is not obvious. In what follows we will investigate
the relation
 between the triad of parameters ($N$, $\varepsilon$, $\Delta t$)
 and how their particular values
 affect the relaxation of the galaxy model.

\section{Numerical tests of galaxy relaxation} \label{conv-stu}
In order to investigate the relaxation and stability against bar
formation for various values of $\varepsilon$, $\Delta t$, and
$N$, the isolated models were evolved up to $t=12.0$, which
corresponds to 3 Gyrs. The main results are summarized in
Table~\ref{galaxy_test}, where the first column gives the model
name, the second the total number of particles, the third the
gravitational softening length, and the fourth contains the
time-step. The rest of the columns represent our results, in the
form of control parameters that are described below. There are
eight series of runs. Models A01-A06 are performed with different
$\varepsilon$ values that were chosen from Fig.~\ref{bar_fig3} at
uniform intervals.
Models A07-A10 and A16-A23 are similar to the first series but for
different $N$. The effect of the time-step is studied in models
A11-A15. The time-steps are decreased by a factor of two, covering
the range of velocities $0.64$--$10$ that satisfy the Courant
condition for $\varepsilon=0.01$. Finally, models A24-A31 are
performed using GADGET with different spline softenings in order
to compare two forms for the force calculation.


\begin{table*}
 \caption{Numerical parameters of galaxy test runs. Tabulated are the model label
 (1), and the input parameters: the number of particles (2), the softening parameter
 $\varepsilon$ (3), and the time step $\Delta t$ (4). The units
 of length and time are $40\ {\mbox {kpc}}$ and $250\ {\mbox {Myrs}}$, respectively.
 As a result of simulations the following control parameters are displayed:
 the relative change of components of the disc velocity dispersions (5-7),
 the disc angular momentum loss (8), the Toomre's Q parameter (9), the Toomre's $X$
 parameter (10), the average value of the distortion parameter (11), the conservation of
 the total energy (12) and the total angular momentum (13) of the system. Finally, we
 mention the code used (14).}
 \label{galaxy_test}
  \begin{tabular}{@{}lccccccccccccc@{}}
  \hline\hline
  (1) & (2) & (3) & (4) & (5) & (6) & (7) & (8) & (9) & (10) & (11) & (12)
  & (13) & (14) \\
  \hline
Model & $N$ & $\varepsilon$ & $\Delta t$ & $\gamma_r$ &
$\gamma_{\varphi}$ & $\gamma_{z}$ & $\frac{\Delta
L_{\mathrm{d}}}{L_{\mathrm{d 0}}}, \%$ & $Q$ & $X_2$ & ${\overline
\eta}$ &
$\frac{\Delta E}{E_0},\%$ & $\frac{\Delta L}{L_0}, \%$ & Code \\
  \hline
 A01 & 40\,960 & 0.025& $1/512$ & 0.782 & 0.452 & 0.815 & 1.5 & 2.0 & 2.0 & 0.065 & 0.01 & 0.13 & GBSPH \\
 A02 & - & 0.020& $1/512$ & 0.670 & 0.623 & 1.115 & 2.1 & 2.1 & 2.3 & 0.033 & 0.01 & 0.05 & -\\
 A03 & - & 0.015& $1/512$ & 0.806 & 0.792 & 1.089 & 2.4 & 2.5 & 1.5 & 0.036 & 0.01 & 0.12 & -\\
 A04 & - & 0.010& $1/512$ & 0.824 & 0.605 & 1.131 & 3.2 & 2.5 & 1.0 & 0.029 & 0.03 & 0.19 & -\\
 A05 & - & 0.005& $1/512$ & 0.863 & 0.676 & 1.249 & 4.5 & 2.8 & 0.9 & 0.026 & 0.04 & 0.19 & -\\
 A06 & - & 0.001& $1/512$ & 0.817 & 0.866 & 1.374 & 6.4 & 3.0 & 0.8 & 0.025 & 2.13 & 0.12 & -\\
\hline
 A07 & 163\,840 & 0.015& $1/512$ & 0.457 & 0.335 & 0.446 & 0.8 & 1.9 & 2.5 & 0.021 & 0.01 & 0.02 & - \\
 A08 & - & 0.010& $1/512$ & 0.541 & 0.443 & 0.532 & 1.0 & 2.2 & 2.5 & 0.019 & 0.01 & 0.02 & -\\
 A09 & - & 0.005& $1/512$ & 0.573 & 0.417 & 0.662 & 1.3 & 2.3 & 2.0 & 0.015 & 0.02 & 0.05 & -\\
 A10 & - & 0.001& $1/512$ & 0.679 & 0.556 & 0.909 & 2.8 & 2.6 & 1.5 & 0.016 & 0.57 & 0.03 & -\\
\hline
 A11 & - & 0.010& $1/64$ & 0.562 & 0.398 & 0.477 & 1.2 & 2.1 & 2.0 & 0.019 & 1.08 & 0.07 & - \\
 A12 & - & 0.010& $1/128$ & 0.527 & 0.397 & 0.466 & 1.0 & 2.1 & 2.0 & 0.017 & 0.04 & 0.05 & -\\
 A13 & - & 0.010& $1/256$ & 0.536 & 0.462 & 0.470 & 1.1 & 2.2 & 2.5 & 0.019 & 0.02 & 0.02 & -\\
 A14 & - & 0.010& $1/512$ & 0.541 & 0.443 & 0.532 & 1.0 & 2.2 & 2.5 & 0.019 & 0.01 & 0.02 & -\\
 A15 & - & 0.010& $1/1024$ & 0.523 & 0.482 & 0.478 & 1.2 & 2.2 & 2.2 & 0.025 & 0.02 & 0.05 & -\\
\hline
 A16 & 491\,520 & 0.010& $1/512$ & 0.347 & 0.269 & 0.195 & 0.6 & 1.9 & 2.6 & 0.016 & 0.01 & 0.04 & - \\
 A17 & - & 0.008& $1/512$ & 0.386 & 0.224 & 0.263 & 0.5 & 2.0 & 2.8 & 0.012 & 0.01 & 0.02 & -\\
 A18 & - & 0.006& $1/512$ & 0.384 & 0.270 & 0.277 & 0.6 & 2.0 & 2.8 & 0.011 & 0.01 & 0.02 & -\\
 A19 & - & 0.004& $1/512$ & 0.340 & 0.277 & 0.316 & 0.7 & 2.1 & 2.8 & 0.009 & 0.02 & 0.02 & -\\
 A20 & - & 0.002& $1/512$ & 0.333 & 0.253 & 0.382 & 0.7 & 2.2 & 2.7 & 0.008 & 0.04 & 0.03 & -\\
 A21 & - & 0.001& $1/512$ & 0.315 & 0.256 & 0.455 & 0.8 & 2.2 & 2.6 & 0.008 & 0.21 & 0.03 & -\\
\hline
 A22 & 655\,360 & 0.008& $1/512$ & 0.320 & 0.247 & 0.155 & 0.4 & 1.9 & 2.8 & 0.010 & 0.02 & 0.01 & - \\
 A23 & - & 0.001& $1/512$ & 0.306 & 0.253 & 0.389 & 0.7 & 2.1 & 2.5 & 0.009 & 0.12 & 0.02 & -\\
\hline
 A24 & 163\,840 & 0.015& $1/512$ & 0.786 & 0.618 & 0.536 & 2.0 & 2.4 & 2.4 & 0.035 & 0.09 & 0.02 & GADGET \\
 A25 & - & 0.010& $1/512$ & 0.861 & 0.648 & 0.566 & 2.6 & 2.5 & 2.1 & 0.030 & 0.12 & 0.06 & -\\
 A26 & - & 0.005& $1/512$ & 0.600 & 0.477 & 0.708 & 2.0 & 2.5 & 1.9 & 0.016 & 0.13 & 0.05 & -\\
 A27 & - & 0.001& $1/512$ & 0.658 & 0.529 & 0.908 & 2.5 & 2.6 & 1.4 & 0.014 & 0.79 & 0.05 & -\\
\hline
 A28 & 655\,360 & 0.008& $1/512$ & 0.674 & 0.505 & 0.271 & 1.3 & 2.1 & 2.6 & 0.018 & 0.21 & 0.57 & - \\
 A29 & - & 0.001& $1/512$ & 0.308 & 0.276 & 0.425 & 0.6 & 2.2 & 2.7 & 0.008 & 0.30 & 0.42 & -\\
\hline
 A30 & 1\,310\,720 & 0.006& $1/512$ & 0.369 & 0.276 & 0.141 & 0.3 & 2.0 & 3.1 & 0.010 & 0.11 & 0.29 & - \\
 A31 & - & 0.001& $1/512$ & 0.267 & 0.229 & 0.312 & 0.5 & 2.1 & 2.8 & 0.008 & 0.13 & 1.11 & -\\
 \hline
\end{tabular}
\end{table*}

The main parameter we use to measure the heating rate of the disc
is the relative change of the velocity dispersion components. The
averaged change in the velocity dispersion is defined over $n$
annulus as:
\begin{eqnarray}
\gamma^2_k=\frac{1}{n}\sum_{i=1}^n\left(\frac{\sigma^2_{k {t_2}}(i)-
\sigma^2_{k {t_1}}(i)}{\sigma^2_{k {t_1}}(i)}\right)^2 ,
\end{eqnarray}
where $\sigma^2_{k {t_2}}(i)$ and $\sigma^2_{k {t_1}}(i)$ are the
$k^{\mathrm{th}}$ component of the velocity dispersions averaged
over the $i^{\mathrm{th}}$ annulus at time $t_2$ and $t_1$,
respectively, and $k=[r, \varphi, z]$. Their values are presented
in columns $5$--$7$ of Table~\ref{galaxy_test}. In absence of the
heating process, these quantities should all be near zero, while
non-zero values indicate an increment in particle velocity
dispersions. The system should be relaxed towards a new
equilibrium, because the initial models are only in approximate
equilibrium and because of the introduction of the softening. For
this we exclude the transient time interval $t<0.5$ Gyrs. We have
evaluated $\sigma^2_k$ at times $t_1=0.5$ Gyrs and $t_2=3$ Gyrs
which gives the accumulative estimation of the disc's anisotropic
``temperature''. Accordingly, the values of the relative change of
the velocity dispersion were averaged over $n=16$ annuli. The
value of the softening that minimizes these quantities will be
taken as optimal.

For all test runs some disc angular momentum loss is observed. The
transfer rate of angular momentum from the disc to the spherical
components is related with the efficiency of dynamical friction.
For a collisionless system in equilibrium, redistribution of
angular momentum is undesired and should be minimized. The eighth
column of the table shows, in percentage, the relative change of
angular momentum with respect to the initial value, ${L_{\mathrm{d
0}}}$.

The stability of an infinitely thin rotating stellar disc to local
axisymmetric perturbations is characterized by the Toomre
 parameter $Q=\sigma_r \kappa
/3.36 G \Sigma$, where $\Sigma$ is the disc surface density
\citep{b57}. Discs with $Q< 1$ are found to be subject to
spontaneous bar formation, otherwise axisymmetric perturbations
are suppressed. Moreover, non-axisymmetric perturbations in a
differentially rotating disc of finite thickness can be suppressed
if the critical radial velocity dispersion is roughly twice the
Toomre critical velocity dispersion \citep{b32}. In addition, the
susceptibility of a thin stellar disc to swing amplification of
the $m=2$ mode is characterized by the $X_m=r\kappa^2/2 m\pi
G\Sigma$ parameter \citep{b72}. The necessary condition to prevent
the model from spontaneous bar growth is $Q\ge 2$ \citep{b52}, and
to prevent swing amplification in galaxies with steep rotation
curves, \citet{b53} found that $X_2>3$ is required. Their values
in our initial models have a minimum of $1.65$ and $3.25$,
respectively, and their final values for each run are shown in
columns $9^{\mathrm{th}}$ and $10^{\mathrm{th}}$ of
Table~\ref{galaxy_test}.

In order to detect the presence of a bar we use the so-called
distortion parameter, defined as \citep*{b39}:
\begin{equation}
\eta = \sqrt{\eta_+^2+\eta_\times^2} ,
\end{equation}
where
\begin{equation}
\label{eta+} \eta_+=\frac{I_{xx}-I_{yy}}{I_{xx}+I_{yy}} , \quad
\eta_\times=\frac{2I_{xy}}{I_{xx}+I_{yy}} ,
\end{equation}
and the moments of inertia are given by
\begin{equation}
I_{ij}=\sum_{k=1}^{N_d} m_k x_k^i x_k^j , \quad i, j=(x,y) .
\end{equation}
This parameter allows us to detect any non-axisymmetric
deformations such as bars in the disc plane. The mean values of
$\eta$ for each run are summarized in the $11^{\mathrm{th}}$
column of the same table. We have found traces of a bar for
$\bar{\eta}\ga 0.02$, which represents a rough threshold.

As criteria to estimate global errors in the force calculation, we
use the conservation of the total energy, $E$, and the total
angular momentum, $L$, whose percentage relative errors are also
displayed in Table~\ref{galaxy_test} (columns 12 and 13,
respectively).

 In general, the models are quite stable and become dynamically
``hot'' ($Q\ge 2$) which should prevent them from spontaneous bar
formation. However, at the same time $X_2$ is reduced, making the
disc susceptible to bar mode amplification. The major disc heating
indeed occurs for $t<0.25$ Gyrs, while the system shifts to a new
equilibrium; later on, the velocity dispersions grow relatively
slowly. From Table~\ref{galaxy_test} a general tendency of $Q$ to
decrease linearly with increasing $\varepsilon$ can be observed
for the Plummer softening.

As can be seen from Table~\ref{galaxy_test}, in models A01-A10,
the variation of $\varepsilon$ affects most of the control
parameters. There is a clear tendency of the parameters
$\gamma_r$, $\gamma_{\phi}$ and $\gamma_z$ to increase with
decreasing $\varepsilon$, caused by the heating process of the
disc. The tendency is pronounced for small $N$, and weakens for
large $N$ (models A16-A23). Indeed, no simple interpretation of
the behaviour of $\gamma_r$ and $\gamma_{\phi}$ on $\varepsilon$
can be drawn for the simulations with higher number of particles
(models A16-A23) and those performed with GADGET (models A24-A31).
However, it is clear that with the increasing $N$ the heating is
reduced.

During the evolution, there are some halo particles that escape
from the system, but their number is relatively small ($< 2 $ per
cent), and for all models the density profile of the spherical
components is well preserved. For small values of $\varepsilon$,
the disc rapidly heats up, and reaches very large values of $Q$
($>2$). In these cases, we observe suppression of a transient
spiral pattern after a few rotation periods, indicating the
presence of dynamical heating of the disc.
 For large $\varepsilon$ the central region
 of the galaxy model with $N=40\,960$ rapidly collapses, and a ring-like
 structure moving outwards is observed during the first few time steps.
 The latter is due to the non-equilibrium evolution of the model caused
 by the lowered potential energy and the strong deviation from the Newtonian
 law.
Regarding the disc vertical structure, it is observed that for
smaller $\varepsilon$, discs become thicker during the evolution.
The two-body relaxation effects in the vertical direction become
important as the vertical structure is being resolved. This makes
the disc hotter and thicker (compare $\gamma_z$ values in
Table~\ref{galaxy_test}).

We allowed some models to run up to $t=8$ Gyrs, at which $\eta$
reaches the value $\sim 0.1$.  Since $t\approx 4$ Gyrs we found a
weak diffuse bar, that appears even for relatively high values for
$Q$ ($>2.0$), that indicates that such high $Q$-values alone
cannot suppress the bar instability. A more massive halo also
prevents the bar formation for a longer time. However,
\citet{Athan03} has recently shown that a live halo can play a
rather destabilizing role. The time at which the bar forms varies
slightly with $N$ and $\varepsilon$, appearing earlier for small
$N$ and large $\varepsilon$ (see values of $\bar \eta$ of the
Table~\ref{galaxy_test}). This confirms findings of \citet{b69},
who showed that the bar emerges later when $N$ is increased, and
for a rigid halo the bar does not form. Large softenings lead to
formation of a bar that is more pronounced, and appears at earlier
times ($t\approx 2$ Gyrs) in spline softening models than in
Plummer's.

To demonstrate that the total energy conservation mainly depends
on the time-step, we vary $\Delta t$ and fix the rest of the
numerical parameters constant. This is done in models A11-A15,
which show that the main change among the control parameters is in
the energy. The deterioration of energy conservation for smaller
gravitational softenings is explained by the errors introduced by
the numerical leap-frog integrator for particles with high
accelerations which are integrated with large time-steps.
 When the softening is small but the
 time-step is sufficiently large, such as for example in model A06,
 a systematic non-conservation of the total energy is observed
 at each time-step.
As shown in models A11-A15, for $\varepsilon=0.01$ and for typical
velocities ${\bar v} \sim 1$ or $\sim 156$ km\,s$^{-1}$, a
time-step of $1/128$ is quite sufficient to obtain energy
conservation of $\sim 0.04$ per cent after $3$ Gyrs of evolution.
On the other hand, comparing the energy conservation between
models with $\varepsilon=0.001$, it can be noted that the increase
of $N$ also improves energy conservation. This is probably due to
the accomplishment of a smoother gravitational field.

For comparison, we have performed several runs using the GADGET
code (models A24-A31). We have decided to skip the runs with
$N=491\,520$ but preformed instead the simulations with
$N=1\,310\,720$. These runs, in general, show higher heating rates
and poorer energy conservation. The higher heating rate of these
models suggests that the Plummer softening makes the system less
collisional than the corresponding spline softening. This is due
to a slower convergence of the Plummer model to the Newtonian
force \citep{theis98}.

The high particle numbers of models A16-A21 and A22-A23 decrease
the overall process of dynamical heating and diminish the
distortions of the disc surface density. By comparing the $\gamma$
values with the same $\varepsilon$ and different number of
particles, we found that augmenting $N$ by four times, the
$\gamma$ values diminish few times. For the given range of
$\varepsilon$ in models A16-A21, the weak variation of most of
their control parameters indicates that they are close to
convergence. Models A22-A23 confirm this convergency. Models
A30-A31 that were performed using GADGET code also show further
reduction of the disc heating and give results comparable to those
of models A22-A23, which were performed with GBSPH code. Note
also, that these models show no signs of a bar.
\begin{figure}
\begin{center}
\includegraphics[width=84mm]{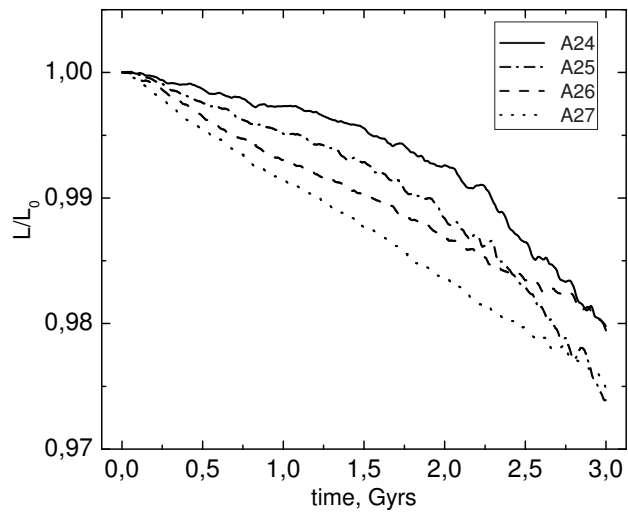}
\caption{Evolution of the disc angular momentum for models
A24-A27, normalized to the initial value.} \label{bar_fig13}
\end{center}
\end{figure}

The effect of the softening clearly manifests itself in the amount
of angular momentum transferred from the disc to the bulge and
halo. The eighth column of Table~\ref{galaxy_test} shows the
dependence of the disc's angular momentum loss on the numerical
parameters. In all runs decreasing the Plummer softening,
increases the disc angular momentum loss. The observed loss rate
is almost constant through a single simulation and is different
for each run. For spline softening (models A24-A31) this process
is slightly different, as can be seen in Fig.~\ref{bar_fig13}. The
loss rate is linear in models A26 and A27 as for models with the
Plummer softening, but in models A24 and A25, as well as in A28,
the bars emerge quite rapidly at $t\approx 2$ Gyrs, which
increases the growth rate. Concerning the total angular momentum,
it can be noted that it is conserved quite well and almost
independent on our chosen numerical parameters.

The numerical tests have shown that for the models in
Table~\ref{galaxy_test}, the softening $\varepsilon=0.001$, for
the considered range of $N$ and $\Delta t$, has the higher heating
rate and the worst total energy conservation, but, at the same
time, smaller non-axisymmetric distortions. This value probably
fulfills a criteria similar to the ones given through equations
(\ref{int_rad}) and (\ref{athanassoula_eps}) for a compound galaxy
model and minimize the initial force errors. However, these models
cannot be accounted as realistic ones (except for the model A31),
but, for comparison, we will use them in some of the collisions
described in the next section. Other extreme cases are those that
use $\varepsilon$ evaluated at the edge of the disc. Despite the
fact that in some models they do not resolve the inner and
vertical structure of the disc, but only its outer part, they show
the smallest heating and a good conservation of $E$ and $L$. These
models are also used in collision runs. From
Table~\ref{galaxy_test} we choose as compromise values for
$\varepsilon_{\mathrm{opt}}$ the ones that correspond to models
A03, A08 and A19. In addition, the models A22-A23 and A28-A29 have
also been taken since they show convergence in the control
parameters.
%
\section{Galaxy collisions} \label{coll}
In this section we describe the effect of the number of particles
and the gravitational softening on the properties of an
interacting system. Galaxy models are constructed with different
$\varepsilon$ and $N$ values and used in galaxy collision
simulations to study the effect of $\varepsilon$ on the formation
of bars and their properties, such as the amplitude and rotational
velocity. In order to reduce the parameters' space, all collision
simulations were performed with the single time-step $\Delta
t=1/256$.

For close galaxy encounters, interaction between particles are
much more important than for an isolated galaxy. The value of
$\varepsilon_{\mathrm{opt}}$ found for an isolated steady state
system is not necessarily optimal for an interacting system where
strong non-linear effects and large density gradients are
developed. Force errors and relaxation processes are drastically
increased by the encounter due to the almost head-on collision of
particles along the shock front. Moreover, the time-step found for
the isolated galaxy models may be inadequate for collision
simulations.

For the simulations, equal galaxies are placed on parabolic
orbits, calculated from the two-body problem with time to
pericentre $t_{\mathrm{p}}=0.75$ Gyrs and pericentric separation
$p=2R_d=32$ kpc.
 The latter is chosen such that at the pericentre the discs are just
 touching each other, and the main perturbation is provided by the
 interaction of the extended haloes. This pericentric separation
 provides enough time for the bars to evolve and, at the same time,
to investigate the merger process. With such orbital parameters,
the galaxies are initially separated by $R_{p}\approx 160$ kpc,
which sets the system with overlapped haloes and thus, it
introduces some initial perturbations. The relatively small
$R_{p}$ is chosen to reduce cumulative numerical errors, where the
preferred duration of interaction should be as small as possible.
The galaxies were relaxed up to $t=0.25$ Gyrs before placed on
their orbits.

We only consider prograde-retrograde collisions that allows us to
investigate, with a single configuration, bar formation processes
in the two possible directions of rotation. Planar collisions
(i.e., disc planes of the galaxies coincide) are chosen, because
 this configuration is the most violent and provides the maximum
 perturbation and maximum transfer of orbital to internal angular
 momentum of the discs.
In the present study we do not pretend to cover the entire orbital
parameters space. The emphasis is rather to investigate the
influence of the numerical parameters on the bar formation for a
given configuration.

 Details of collision models and conservation of total energy and
 angular momentum are summarized in Table~\ref{collisions}.
 Animations of some collision simulations are available at
 {\it www.astro.inin.mx/ruslan/tidal\_bars}.

\begin{table}
  \caption{Parameters of collisions. The units
 of time and length are $250\ {\mbox {Myrs}}$ and $40\ {\mbox {kpc}}$, respectively.}
  \label{collisions}
  \begin{tabular}{@{}lcrcrcc@{}}
  \hline\hline
   Model & $N$ & $\Delta t$ & $\varepsilon$ & $\frac{\Delta E}{E_0}$, \%
 & $\frac{\Delta L}{L_0}$, \%  & Code \\
 \hline
 B01 & 40\,960 & 1/256 & 0.001 & 14.53 & 0.06 & GBSPH \\
 B02 &  -   & 1/256 & 0.015 & 0.09 & 0.10 & - \\
 B03 &  -   & 1/256 & 0.025 & 0.14 & 0.40 & - \\
\hline
 B04 & 163\,840 & 1/256 & 0.001 & 5.40 & 0.05 & - \\
 B05 &  -   & 1/256 & 0.010 & 0.11 & 0.10 & - \\
 B06 &  -   & 1/256 & 0.015 & 0.10 & 0.16 & - \\
\hline
 B07 & 491\,520 & 1/256 & 0.001 & 2.17 & 0.07 & - \\
 B08 &  -   & 1/256 & 0.004 & 0.10 & 0.16 & - \\
 B09 &  -   & 1/256 & 0.010 & 0.09 & 0.11 & - \\
\hline
 B10 & 655\,360 & 1/256 & 0.001 & 1.61 & 0.06 & - \\
 B11 &  -   & 1/256 & 0.008 & 0.07 & 0.08 & - \\
\hline
 B12 & 163\,840 & 1/256 & 0.001 & 7.0 & 4.98 & GADGET \\
 B13 &  -   & 1/256 & 0.010 & 0.33 & 5.87 & - \\
 B14 &  -   & 1/256 & 0.015 & 0.27 & 4.98 & - \\
\hline
 B15 & 655\,360 & 1/256 & 0.001 & 2.17 & 2.94 & - \\
 B16 &  -   & 1/256 & 0.008 & 0.11 & 2.33 & \\
\hline
\end{tabular}
\end{table}

In order to perform a reliable analysis of the bar
characteristics, its centre of mass should be well defined. The
disc particles that are ejected to large distances can spuriously
affect the parameters under study, and therefore these particles
are excluded from the analysis. We then calculate the centre of
the disc as follows: First we compute the centre of mass using all
the particles of the disc. We then recalculate a new centre of
mass with only the disc particles that are located inside $2 R_d$
with respect to the firstly calculated centre of mass. The centre
of mass found in such a way gives roughly the position of the
centre of mass of the disc with respect to which the bar
parameters are further computed. However, this procedure fails
during the last stage of the collision process when both systems
are found inside the same radius.

To characterize quantitatively the bar, we use two criteria. The
first is given through the distortion parameter $\eta$ defined
above. The second is to compute the harmonic amplitudes of the
disc mass distribution in the equatorial plane, which indicate the
presence of any non-axisymmetric deviation.
 The normalized coefficients are \citep{b38}
\begin{equation}
A_m = \frac{1}{n} \sum_{j=1}^n \exp(i m\theta_j),
\end{equation}
where $n$ is the number of particles within the radius of the disc
$R_{\mathrm{d}}$, and $\theta_j$ is the polar angle of the
$j^{th}$ particle with respect to the centre of mass. Since we are
mostly interested in bisymmetric perturbations, we set $m=2$, but
we also investigate amplitudes up to $m=8$ to compare among the
harmonics. We have found that the $\eta$ and $|A_2|$ bar strengths
criteria give similar results. However, stronger fluctuations are
observed using $\eta$, and in what follows we will only refer to
$|A_2|$.
\begin{figure}
\begin{center}
\includegraphics[width=84mm]{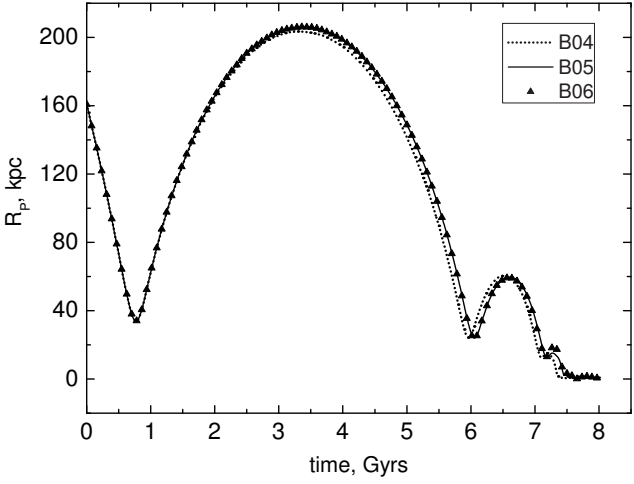}
\caption{The separation between the centres of the two discs as a
function of time for models B04-B06.} \label{bar_fig4}
\end{center}
\end{figure}
\begin{figure}
\begin{center}
\includegraphics[width=84mm]{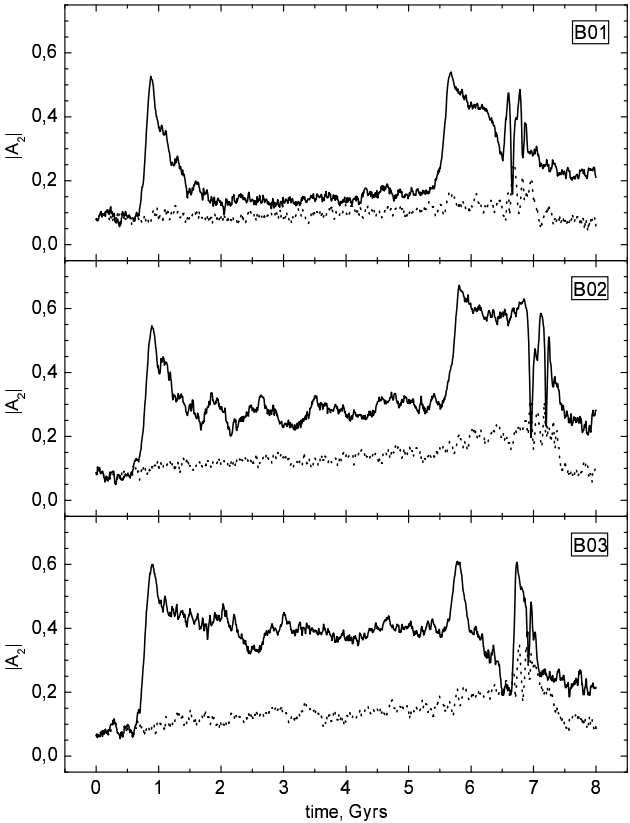}
\caption{The evolution of the amplitude of the second harmonic for
models B01-B03. The solid line corresponds to the first galaxy
(prograde orbit) whereas the dotted line corresponds to the second
galaxy (retrograde orbit).} \label{bar_fig5}
\end{center}
\end{figure}
Some morphological characteristics of the bar can be obtained from
projected density contours and projections of particles on the
orthogonal planes.

 An important quantity that characterizes the bar is its
 rotational velocity $\Omega$, that is given by the rate of change of the
 angle between the coordinate axis and the principal axis of the
 moment of inertia of the bar.
 We divide the plane of the disc into $16$ concentric cylinders for
 which the eigenvalue of the moment of inertia tensor $I_{xx}$ is
 calculated.
 Then we iteratively rotate the coordinate system with an angular
 step of $2^{\circ}$ until the maximum value of $I_{xx}$ is reached
 for each annuli.
 The corresponding averaged polar angle
 of rotation $\phi$ with respect to the original coordinate system is
 taken as the phase of the bar. The procedure is repeated for each
 snapshot file, and
 $\Omega$ is found as $d\phi/dt$, where $dt=1/32$ is the time
 interval between snapshots.
 To correctly identify the bar major axis,
 we impose the restriction on the minimum value $\eta_+=0.15$, which
excludes the time intervals when the bar is not present and
excludes also the central annulus, which is almost axisymmetric.
The second restriction consists to discard annuli with a phase
difference of more that $10^{\circ}$ from the averaged phase of
the inner annuli. This procedure excludes the annuli where spiral
arms and the ring are located.
\begin{figure}
\begin{center}
\includegraphics[width=84mm]{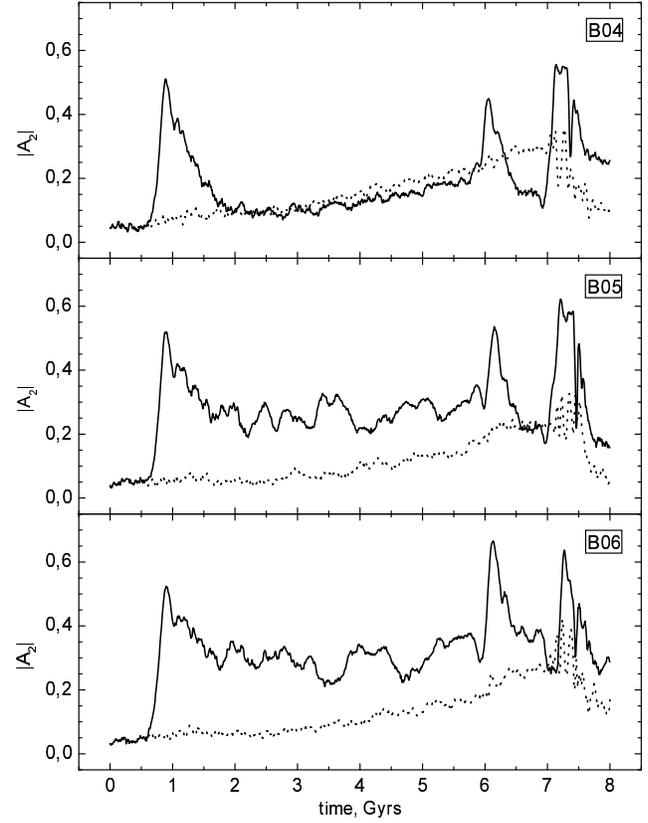}
\caption{The evolution of the amplitude of the second harmonic for
models B04-B06. The correspondence of curves is the same as for
the previous figure.} \label{bar_fig6}
\end{center}
\end{figure}
\begin{figure}
\begin{center}
\includegraphics[width=84mm]{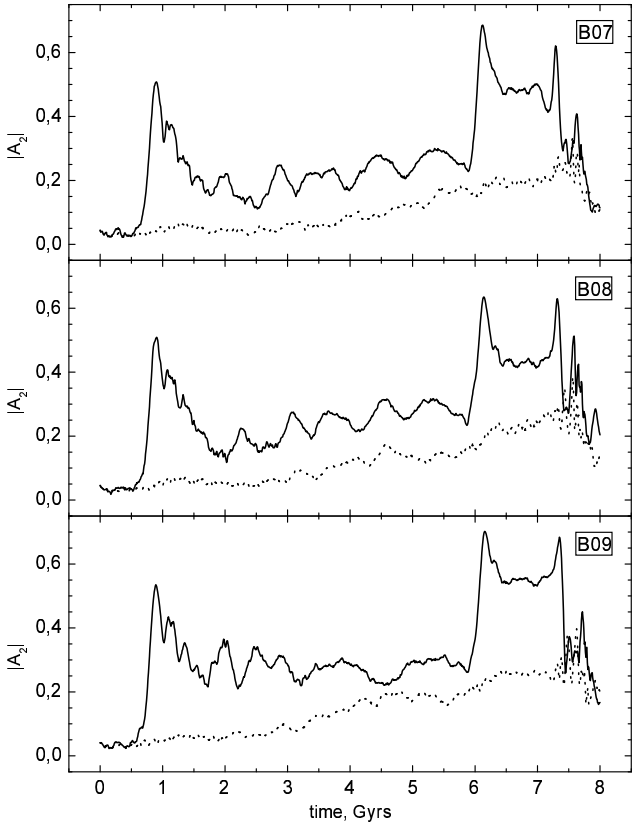}
\caption{The evolution of the amplitude of the second harmonic for
models B07-B09. The correspondence of curves is the same as for
Fig.~\ref{bar_fig5}.} \label{bar_fig7}
\end{center}
\end{figure}
\begin{figure}
\begin{center}
\includegraphics[width=84mm]{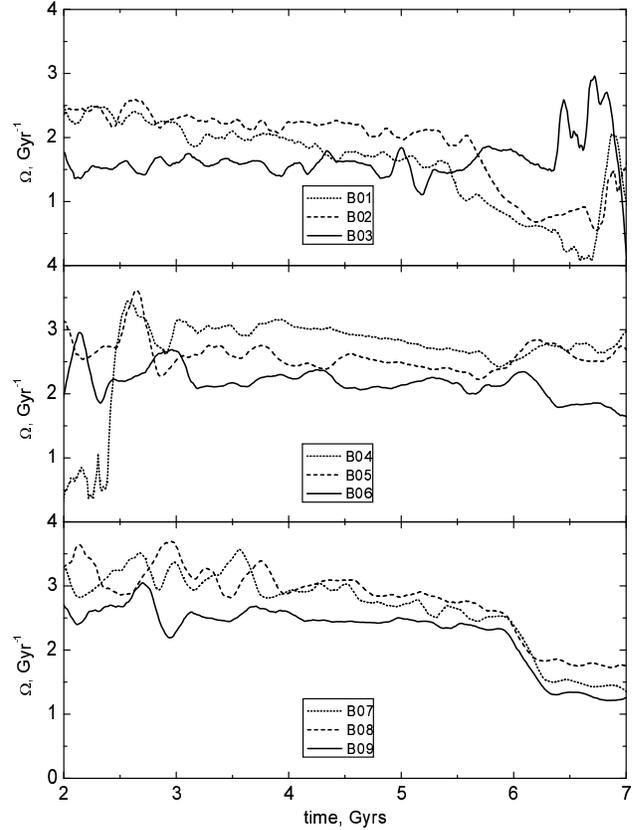}
\caption{The evolution of bar angular velocity for collision runs
B01-B09. Shown only $\Omega$ of the first (direct) disc.}
\label{bar_fig12}
\end{center}
\end{figure}
\begin{figure}
\begin{center}
\includegraphics[width=84mm]{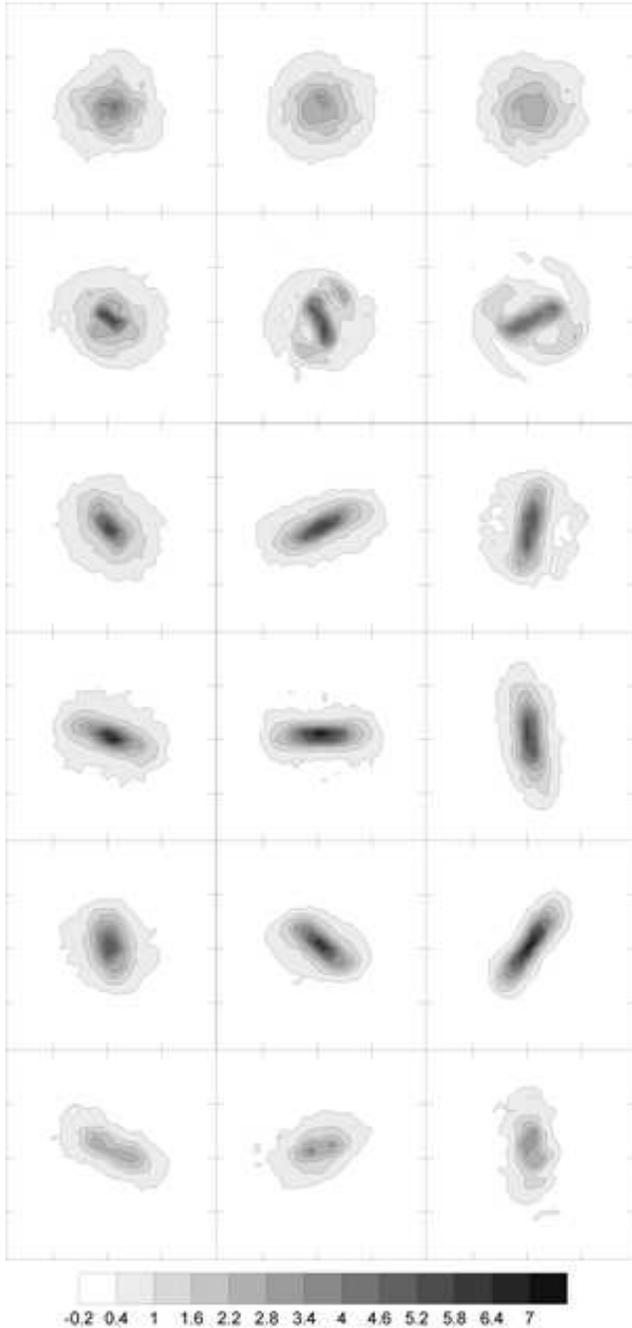}
\caption{Projected density contours of the first (prograde) disc
in units of $137.5\ {\mbox M}_{\sun}/{\mbox {pc}}^2$. Columns from
left to right correspond to models B04, B05 and B06. The boxes are
$32\times 32$ kpc long. Contours are shown for times (panels from
top to bottom) 0, 1.5, 3.5, 5.5, 6.5 and 8 Gyrs.}
\label{bar_fig10}
\end{center}
\end{figure}
\begin{figure}
\begin{center}
\includegraphics[width=84mm]{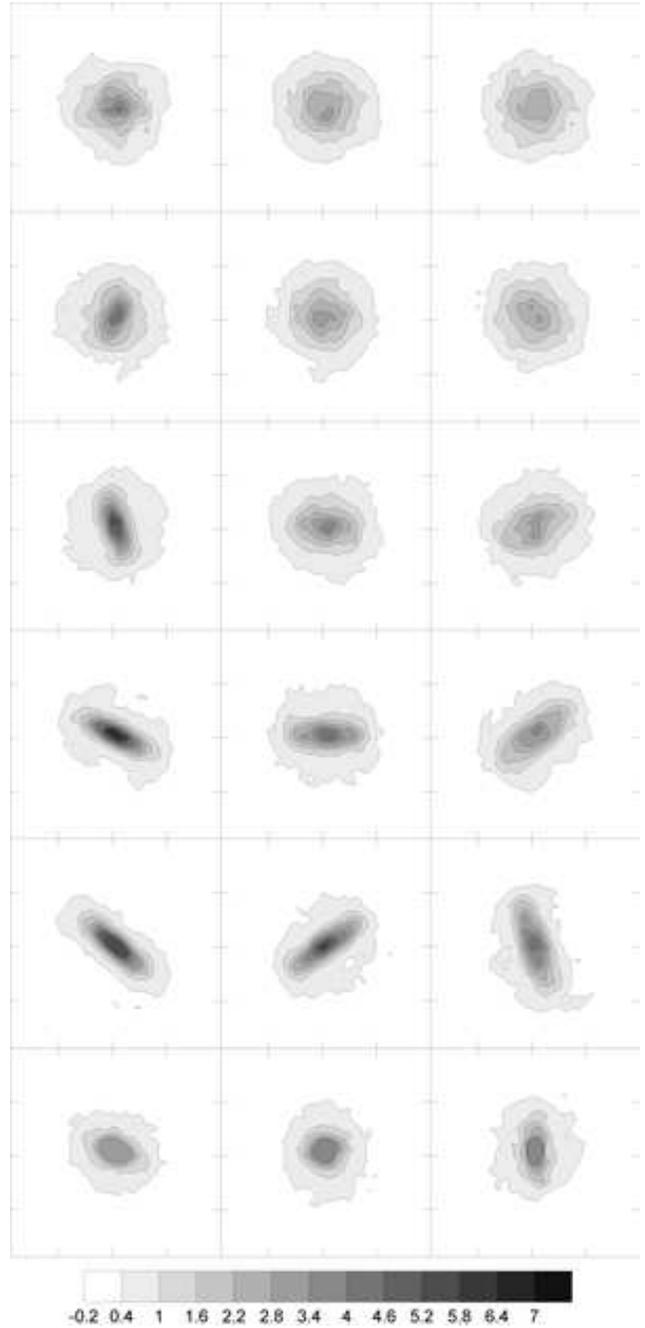}
\caption{Projected density contours of the second (retrograde)
disc. Columns from left to right correspond to models B04, B05 and
B06 respectively. All scales are the same as in the previous
figure.} \label{bar_fig11}
\end{center}
\end{figure}
\begin{figure}
\begin{center}
\includegraphics[width=84mm]{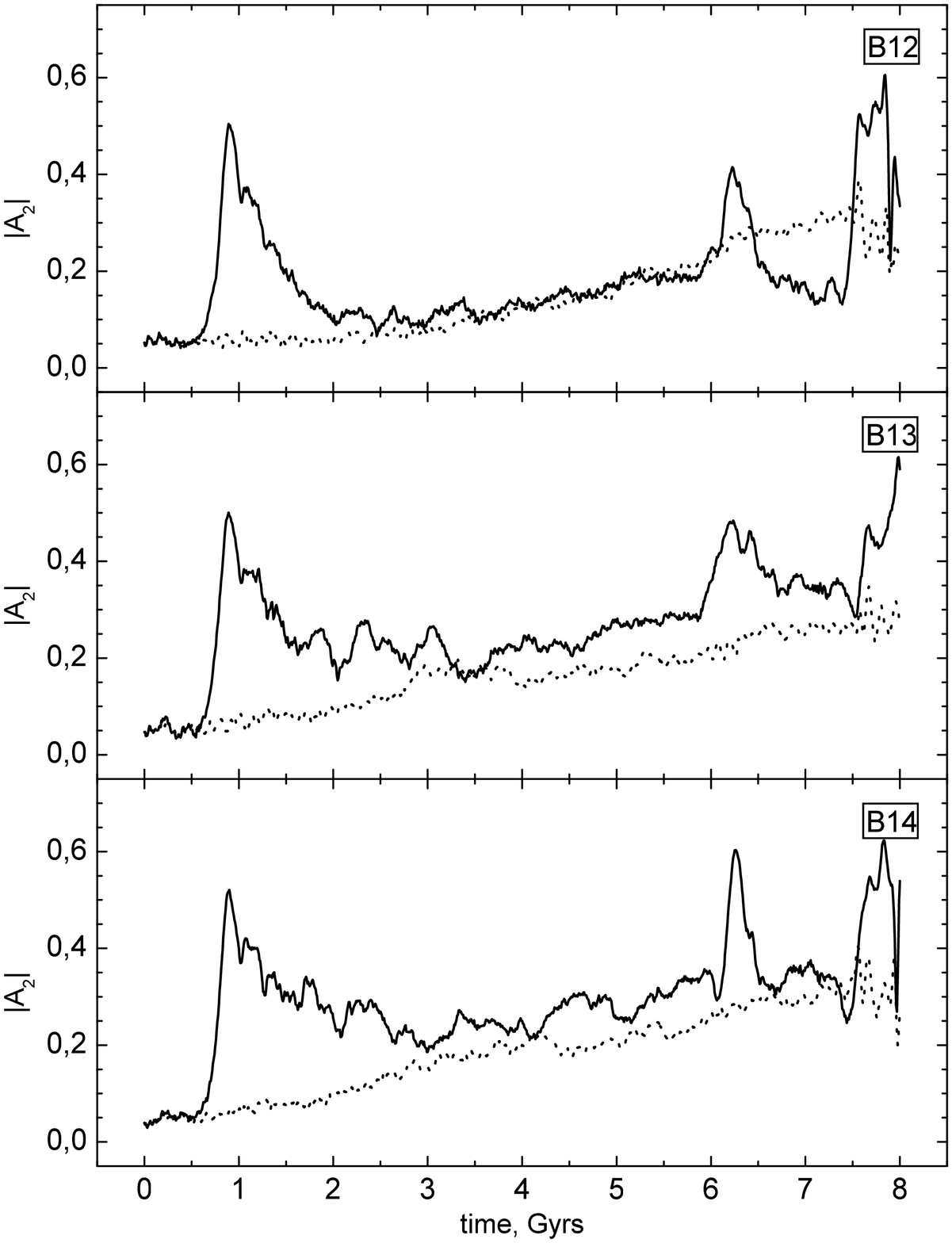}
\caption{The evolution of the amplitude of the second harmonic for
models B12-B14. The correspondence of curves is the same as in the
Fig.~\ref{bar_fig5}.} \label{bar_fig8}
\end{center}
\end{figure}

For all galaxy collision models, the first encounter occurs at
time $t\approx 0.8$ Gyrs and the second at $t\approx 6$ Gyrs. The
tidal interaction induces a bipolar perturbation that develops
spiral arms in the first galaxy which is in prograde orbit, while
the second galaxy does not show any significant deformation. The
subsequent evolution is different for each model as will be
described below.

Fig.~\ref{bar_fig4} shows the separation between the centres of
the two discs as a function of time for models B04-B06. It can be
noted that in model B04, after the first encounter, the discs have
a maximum separation distance at $t\approx 3.5$ Gyrs, which is
less than those of models B05 and B06 due to a smaller softening
of the force. In addition, the whole collision process is slightly
faster for model B04, and slower for B05 and B06. By analysing the
curves for runs B01-B03 (not shown here) we have noted that they
have a similar shape, and that by increasing $N$ (collisions
B07-B09) the difference tends to vanish.

Models B01-B09 were performed with the GBSPH code with three
different values of $N$ and $\varepsilon$. We first describe
collision runs B01-B03 which were performed with $N=40\,960$
particles per galaxy. As can be seen from Table~\ref{collisions},
the run B01 shows bad conservation of the total energy due to
integration errors caused by inadequate time-step for such small
$\varepsilon$. But at the same time, there is a good angular
momentum conservation. The evolution of the bar amplitude for
these runs is shown in Fig.~\ref{bar_fig5} for both prograde
(solid line) and retrograde (dotted line) discs. There are also
small contributions from other harmonics, mostly from even ones.

 After the first encounter, transient spiral
 arms are formed in the prograde galaxy which then
 become more and more tightly wound until they form a disc again. At
 the same time, a small diffuse bar emerges which maintains an
 oval shape of apparent length $l \sim 10$ kpc until the second collision,
 after which the bar is amplified to length $l \sim 16$ kpc and
 develops a butterfly shape when
viewed from the disc's plane.

In model B03 after the first passage, transient open spiral
pattern is generated in the first galaxy, and at the same time, a
strong bar of length $l \sim 20$ kpc appears. Note that this is
the simulation with the longest bar. The spiral arms begin to wind
around the bar, forming a ring around it which is then slowly
absorbed by the bar. The bar maintains its length until the second
close approach, after which it is partially destroyed and reduced
to $l \sim 16$ kpc.

The second galaxy that is in retrograde motion shows a slowly
growing bar mode, which is amplified by the second encounter (see
Figs.~\ref{bar_fig5}-\ref{bar_fig7}). The barred galaxies collide
again and finish in a disc-like remnant. The initial growing mode
of the second galaxy that becomes notable after $t\approx 4$ Gyrs
is similar to the one in the isolated evolution, described in
sec.~\ref{conv-stu}. Therefore, it seems not to be caused by the
tidal interaction, but instead by the instability of the galaxy
model.
 This is observed in all collision models presented in
Table~\ref{collisions}. The instability growth rate is slow and by
the time of the first encounter the bar has not yet been formed.
However, the evolution of the bar amplitude after $t\approx 6$
Gyrs may be influenced by both effects.

The behaviour of the B02 collision run, is similar to the B03,
except that the bar is shorter ($l\sim 16$ kpc). After the second
collision, the bar is strongly amplified, because at the
pericenter it becomes aligned with a diffuse bar of the second
galaxy. The amplitude of the second harmonics of the bar reaches
the value $\sim 0.6$, but apparent length of the bar does not
change, obtaining rather a butterfly shape.

The models B04-B06 are performed with $N=163\,840$ particles per
galaxy and show a behaviour similar to models B01-B03 but with
richer details in the inner structure. For example, a $\S$- shaped
structure surrounding the bar was present in these runs, but it
was not observed in simulations with $N=40\,960$ particles. The
notable tidal perturbation decay is also observed in model B04,
but the difference between the models B05 and B06 is not so strong
(Fig.~\ref{bar_fig6}). The models B07-B09 are performed with
higher $N$ and show a more detailed inner structure, but only a
weak dependence of bar strength on $\varepsilon$ (see
Fig.~\ref{bar_fig7}).

Similar differences are observed in the rotational velocities of
the bars. The evolution of bar angular velocities for models
B01-B09 are plotted in Fig.~\ref{bar_fig12}. As it can be noted
form the upper panel, the magnitude of $\Omega$ differs
significantly among the runs. However, with the increase of $N$
these differences are reduced, and the shape of the curves becomes
similar (models B07-B09). A clear evidence for correlation between
the bar strength and its angular velocity can be seen when
Figs.~\ref{bar_fig5}-\ref{bar_fig12} are compared. For example,
after the second encounter the decay of $\Omega$, observed in
models B07-B09 occurs when the bar, due to alignment with the
second bar, is amplified. This reduction can be associated with
the angular momentum exchange which amplifies the bar and reduces
its pattern speed, in accordance with results of \citet{b74}.

The softening affects not only the shape of the bars, but also its
internal structure. Figs.~\ref{bar_fig10} and \ref{bar_fig11}
present the contours of the projected density for models B04-B06
at different $t$. From the upper panels it can be seen that due to
different $\varepsilon$, the initial discs have different central
densities, which may be responsible for their subsequent
evolution. The rest of the panels show that bars with smaller
$\varepsilon$ are rounder, denser and smaller than those with
larger $\varepsilon$.

Summarizing, the results of models B01-B09 have shown that the
decrease of $\varepsilon$ damps perturbations in the prograde
disc. For the second galaxy which is retrograde, the amplitude
$|A_2|$ slowly grows with time, and the growth rate slightly
depends on the chosen $\varepsilon$.
Figures~\ref{bar_fig5}-\ref{bar_fig7} indicate that the initial
perturbation has almost the same peak value of $|A_2|\ga 0.5$, but
further decay of tidal response depends on $\varepsilon$ and $N$.

For comparison, we have performed collision simulations using a
parallel version of the GADGET code (runs B12-B16). As can be seen
from Table~\ref{collisions}, the total angular momentum is poorly
conserved. After the first encounter, this non-conservation shifts
the galaxies from the orbital plane, which is more notable for
small $N$ (runs B12-B14). As a consequence, further encounters are
not planar and the merger process takes longer. Despite this, the
first stage of the bar formation can be compared with previous
runs. Fig.~\ref{bar_fig8} shows the evolution of the amplitude of
the second harmonic for collisions B12-B14, which is similar to
models B04-B06 except that the amplitude of the bar is a bit
smaller and a stronger bar develops in the second disc. The
collision simulations performed using GBSPH and GADGET codes for
$N=655\,360$ show no significant differences in $|A_2|$ for the
same $\varepsilon$; these curves are not shown.

\begin{figure}
\begin{center}
\includegraphics[width=84mm]{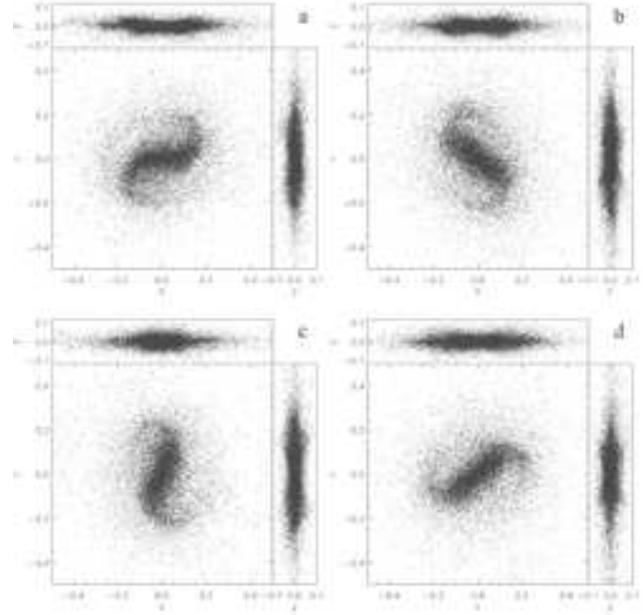}
\caption{Projection of particles of the first disc for run B06 for
times: a) $3.125$, b) $3.875$, c) $4.375$ and d) $5.125$ Gyrs. For
face-on and edge-on projections the boxes are $40\times 40$ kpc
and $8\times 40$ kpc long, respectively. The bar always rotates
clockwise, but its arms wiggle from trailing to leading and
vice-versa.} \label{bar_fig9}
\end{center}
\end{figure}

Finally, we will describe some peculiar characteristics of
collision and merger processes for different softenings. A general
interesting feature is observed between the first and the second
collisions. The two large spiral arms that are formed just after
the first collision in the prograde galaxy become more and more
tightly wound, and form a ring-like structure that surrounds the
diffuse bar. This ring is then slowly destroyed accreting into the
bar, and short transient arms are formed at the extremes of the
bar (Fig.~\ref{bar_fig9}). These arms change from trailing to
leading and vice-versa with a period $T_{\mathrm{arm}}\approx
0.675$ Gyrs (model B06) though the direction of rotation of the
bar does not change. We refer to this dynamical feature as the
{\it wiggle effect}. As most particles are absorbed by the
remaining pronounced bar, the wiggle slowly vanishes but does not
completely disappear and is sustained until the second collision.
The wiggle is observed in runs where a clear bar appears and in
all runs with $N=491\,520$. In the same figure, shown aside, a
classical bar buckling leading to the box/peanut shape may be
seen.

In our orbital configuration, after the first encounter up to $5$
per cent of the particles of the first disc leave the initial
radius, and up to $10$ per cent after the second. The second disc
looses only a few particles during the collision. The particle
loss rate is almost independent on the numerical parameters and in
each case the merger remnant is composed of roughly $70$ and $85$
per cent of particles of the first and second discs, respectively.
In runs B01-B06 the remnants have elongated orbits due to
particles of the prograde disc, whereas particles of the
retrograde disc form a symmetric constant density core
(Figures~\ref{bar_fig10}, \ref{bar_fig11} bottom panel). The
mergers with large softenings finish in a disc-like remnant,
whereas for small softenings, the remnant has rather an elliptical
structure.
%
\section{Discussion} \label{discu}
The aim of the convergence study was to investigate how the
numerical parameters influence the relaxation and stability of a
galaxy model against bar formation. We first discussed how the
parameters $N$, $\varepsilon$ and $\Delta t$ can be restricted,
and for a particular galaxy model obtained valid ranges for
$\epsilon$ and $\Delta t$.
 For a galaxy model represented by a given number of
particles, the local mean interparticle separation in the plane of
the disc can be easily estimated. Although the gravitational
softening parameter can be chosen to be equal to any particular
value of $\lambda$ within the established range, one still has to
make a compromise between force accuracy and relaxation. But it
seems that there are no universal choices, and the selection of a
particular $\varepsilon_{\mathrm{opt}}$ value for for a whole
system depends on dynamical aspects one wants to study
\citep{Romeo98}. In this context we wish to minimize the
relaxation effects that are responsible for drifting of a system
from its initial condition. The convergence study performed in
section~\ref{conv-stu} showed that relaxation processes are
minimized for $\varepsilon$ equal to the maximum
$\lambda(R_{\mathrm{d}})$, which is evaluated at the edge of the
disc. This $\varepsilon$ gives low heating and good energy
conservation, but it neither resolves the vertical nor the radial
disc mass distributions. On the other hand, the minimum
$\varepsilon=\lambda(0)$ is unacceptable in simulations because of
the high heating rate, and because it demands a too small
time-step to achieve acceptable energy conservation. In
Table~\ref{galaxy_test} we have presented a series of models with
gravitational softenings chosen within the established range which
is defined by $N$. Fig.~\ref{bar_fig3} indicates that the
$\lambda$ range decreases with increasing $N$.

We have found that for our galaxy models the characteristics of
the resulting systems tend to converge for $N\ga 491\,520$.
Namely, the reduction of the range of $\lambda$ by increasing $N$
also reduces the variation of control parameters. The parameters
$\gamma_z$, $Q$ and the disc angular momentum loss rate, used to
measure the disc heating, proved to be quite informative. The disc
angular momentum loss rate gives the efficiency of the dynamical
friction between the disc and the halo, whereas the increase of
the vertical velocity dispersion, observed in all runs and for
both softening methods, measures the thickening of the disc.
However, it seems that those quantities are related. The dynamical
friction is enhanced by increasing the thickness of the disc, and
this, in turn, increases the angular momentum loss rate. Indeed,
the change in the slope of angular momentum loss of the disc
serves as a good indicator to identify the bar's emergence. As can
be seen from Table~\ref{galaxy_test}, simulations with $491\,520$
particles greatly reduce the disc angular momentum loss, which is
less than $1$ per cent within $3$ Gyrs, and only weakly depends on
$\varepsilon$. The disc thickening is also reduced to about three
times, when compared to runs with $40\,960$ particles.
 Simulations with $N=1\,310\,720$ show that the heating process
is weakly affected by $\varepsilon$ in comparison with the
simulations with smaller particle numbers for the corresponding
ranges of $\lambda$. From the same table it can be noted that
final values of Toomre's parameter $Q$ for the Plummer softening
show approximately a linear dependence on $\varepsilon$.

The heating of the disc can be reduced by increasing either the
softening parameter or the number of particles. However,
increasing $\varepsilon$ within the allowed range reduces the
heating in a slower way than increasing $N$ (compare the values of
$\gamma$ in the Table~\ref{galaxy_test}). This reaffirms the
common belief that the increasing of the number of particles lets
to effectively decrease the numerical effects and at the same time
to increase the resolution. To obtain the minimum $N$ that
resolves the disc vertical structure, it is necessary that
$\varepsilon\la z_0$. The criterion of the interaction radius,
given through equation (\ref{int_rad}), showed to be inadequate
because it gives too small $\varepsilon$ values. On the other
hand, the criterion given through equation (\ref{epsilon}) would
give a minimum $\varepsilon$, which is approximately $0.3$ times
the mean interparticle separation, estimated within the half-mass
radius. The interaction force in this case will be well
represented and the bar will appear later, but the system will be
heated.

On the other hand, the time-step, which is proportional to the
gravitational softening parameter, defines the accuracy of the
integration of the orbits and is the main factor responsible for
energy conservation. The criterion for the time step $\Delta t \la
\varepsilon/{\bar v}$ showed to be satisfactory, at least in
simulations performed with the Plummer softening (GBSPH). For a
fixed $N$ the energy conservation deteriorates when $\varepsilon$
is reduced. If $N$ is increased and $\varepsilon$ is held fixed,
an improvement in energy conservation is observed. These
parameters have been chosen to obtain an energy conservation of
better than $0.1$ per cent for acceptable models, see
Table~\ref{galaxy_test}.

In agreement with \citet{theis98}, the Plummer softening is less
collisional and has a larger relaxation time than the spline
softening, even when the correspondence is made by equating their
potentials depths. He shows, however, that the same relaxation
time can be obtained with both the spline softening and the
Plummer softening for $h\ga 3\varepsilon$. As a consequence, in
order to maintain the same degree of relaxation and to resolve the
vertical structure of the disc when the spline softening is used,
many more particles are required than for the Plummer model. On
the other hand, the Plummer softening, in comparison with the
spline softening, has a stronger stabilizing effect
\citep{Romeo97, Romeo98}. This may lead, as we have seen, to a
late bar emergence. Further investigations of galaxy model
stability using adaptive softening kernels would be worthwhile
\citep{Dehnen01}.

This work was in part motivated to explain the difference in the
characteristics of the resulting bars in various authors galaxy
collision simulations. For example, works of \citet{b10, b8} show
very strong and prominent bars that formed after the first
encounter. On the other hand, in calculations made by
\citet*{Dubinski96} the apparent bars are not observed. Although
the difference may be due to different galaxy models, we decided
to verify the numerical aspects of the problem. For example, both
galaxy models have roughly the same number of particles and the
time-step, but $\varepsilon$ used by \citet{b8} is six times
larger than Dubinski's.

Our collision simulations of tidally triggered bar formation
showed that the bar properties depend on the selected
gravitational softening parameter. Within the studied ranges of
$\varepsilon$ and $N$, tidal bars are shorter for smaller
$\varepsilon$. Indeed, for sufficiently small $\varepsilon$ and
$N$, tidal distortions can be suppressed. For example, we
performed some additional collision simulations with
$\varepsilon=0.0005$ and $\Delta t=1/1024$ for $N=40\,960$ and
$N=163\,840$, where we were able to completely suppress the bar
formation, and obtained small dense discs. Such models are however
highly collisional and thus cannot be considered as correct. On
the other hand, the bar's pattern speed and the bar's length vary
weakly with $\varepsilon$ and tend to converge to a single value
already for $N\ga 491\,520$. Certainly, simulations with several
million particles will approach further convergence and will show
many more bar details \citep{ONeill03}, but an adequate
$\varepsilon$ and $\Delta t$ still have to be chosen.

Confirming the results of \citet{b74}, we observe a correlation
between the bar's length and the pattern speed. We found also,
that in simulations with large $\varepsilon=\lambda(R_d)$, the
bars roughly conserve their length and pattern speed at least
within $3$ Gyrs, which is time between the first and the second
collisions when a clear bar is observed. However, for smaller
$\varepsilon$ the bar length increases and the rotational velocity
reduces.

The collision simulations have shown that a colder disc (large
$\varepsilon$) forms an open spiral pattern, corresponding to a
late type galaxy, whereas the spiral arms in a hot disc (small
$\varepsilon$) are tightly wound. This is explained by the
dispersion relation for WKB long-branch waves if the disc's
perturbation is weak \citep{b72,b46}. Another possible explanation
of tidal response differences may lie in the increased central
density of relaxed models for small $\varepsilon$, which may
create the inner Lindblad resonance that prevents the feedback of
swing amplification mechanism \citep{Norman96}.

The wiggle effect observed in our simulations could imply that
some real barred galaxies do not necessarily have trailing arms. A
detailed study of this effect, including gas dynamics, would be
worth.
%
\section{Conclusions}\label{conclusions}

In this work we have investigated the influence of the triad of
numerical parameters ($N$, $\varepsilon$, $\Delta t$) on the
properties of an equilibrium isolated galaxy model and the
dynamics of two interacting galaxies. The main results for the
galaxy equilibrium models can be summarized as follows:
\begin{enumerate}

\item The transfer of the disc angular momentum to the bulge
and the halo depends on $\varepsilon$, $N$, and type of softening.

\item The convergence study indicates that for $N=491\,520$ and
$\varepsilon=[0.001-0.01]$, the range of final values of $Q$
changes only by $14$ per cent, while for smaller $N$, the
$\varepsilon$ and $Q$ range broadens. Further decrease of
numerical heating is observed for larger $N$ simulations, namely,
for $N = 1\,310\,720$, $Q$ only varies $5$ per cent.

\item The final value of the Toomre stability parameter decreases
linearly with the increasing of the Plummer softening.

\item For the same values of $N$, $\Delta t$ and
$\varepsilon$, the spline softened simulations produce bars
earlier than those performed with Plummer softening.
\end{enumerate}

For the galaxy collision models it may be concluded that:
\begin{enumerate}

\item The susceptibility of galaxy models to tidal bar formation
and its dissolution depend strongly on the chosen numerical
parameters. Therefore, carefully chosen parameters have to be
taken for specific simulations.

\item For sufficiently small $N$, the size of the bar formed
during an encounter, may become too long or even not appear
depending on the value of $\varepsilon$. With the increase of
 $N$ artificial effect of the softening is reduced.

\item The short spiral arms that are formed at the ends of the bar
show a periodic change from trailing to leading and vice-versa --
the wiggle.

\item The spline softening gives smaller bars than in case of the Plummer
softening even when they have the same potential depth.

\item The properties of the merger remnant are also affected by
the softening. The remnants of cold systems maintain their barred
thick disc shape, whereas those of hot systems produce rather an
ellipsoid.
\end{enumerate}

\begin{acknowledgements}
This work has been partially supported by the Mexican Consejo
Nacional de Ciencia y Tecnolog\'{\i}a (CONACyT) under contracts
U43534-R, 44917-F and J200.476/2004, and the DFG and DAAD of
Germany. RFG also acknowledges the Ministry of Foreign Affairs of
Mexico - ``Secretaria de Relaciones Exteriores" for financial
support. Some of the runs were performed at the Altix 370 computer
facility kindly provided by SGI.
\end{acknowledgements}

We also thank the referee for very useful comments that improved
the presentation of the paper.



\label{lastpage}

\end{document}